\documentclass[twocolumn]{aastex62}

\newcommand{\RNum}[1]{\uppercase\expandafter{\romannumeral #1\relax}}
\newcommand{\bef}{\begin{figure}}      
\newcommand{\eef}{\end{figure}}      
\newcommand{\bea}{\begin{eqnarray}}    
\newcommand{\eea}{\end{eqnarray}}      
\newcommand{\be}{\begin{equation}}      
\newcommand{\ee}{\end{equation}}  
\newcommand\HI{$\textrm{H}\scriptstyle\mathrm{I}$}

\graphicspath{{./}{figures/}}

\received{???}
\revised{???}
\accepted{???}
\shorttitle{Constraining the Geometry of Galactic Dark Matter with Gaia DR3}
\shortauthors{Sylos Labini \& Capuzzo Dolcetta}
\usepackage{amsmath}

\begin{document}

\title{Constraining the Geometry of Galactic Dark Matter with Gaia Data Release 3}
\author{Francesco  Sylos Labini}
\affil{Centro  Ricerche Enrico Fermi, 00184 Rome, Italia}
\author{Roberto Capuzzo-Dolcetta}
\affil{Dipartimento di Fisica, Universit\`a ``Sapienza'', 00185 Roma, Italia}
\affil{Centro  Ricerche Enrico Fermi, 00184 Rome, Italia}

\correspondingauthor{FSL}
\email{sylos@cref.it}

\begin{abstract}
We derive both the mid-plane and off-plane rotation curves, $v_c(R,z)$, and the vertical acceleration, $a_z(R,z)$, of the Milky Way (MW) using \textit{Gaia}~DR3 data over the ranges of vertical heights $z \in (-2,2)\,$ kpc and galactocentric distances $R \in (8.5,14)$ kpc where 
 the velocity components are determined with high precision, {    i.e., with an error $< 5\%$}.  In contrast, the vertical acceleration $a_z(R,z)$ is dominated by model-dependent systematics, with uncertainties of up to $\sim 20\%$.
This level of accuracy allows us to place stringent constraints on the geometry of the MW's dark matter (DM) distribution, as the vertical gradients of the gravitational potential attain their maximum within this range of radial and vertical distances corresponding to the characteristic scales of the disk. 
We find that models including the observed stellar components together with a spherical DM halo fail to reproduce both the pronounced variation of $v_c(R,z)$ with height and the observed behavior of $a_z(R,z)$. 
In particular, spherical halos with a scale radius of $r_s \sim 15$ kpc contribute negligibly to the off-plane rotation curve and vertical acceleration in the inner disk, leaving these features primarily determined by the stellar mass distribution. 
Conversely, models in which DM is confined to a flattened, disk-like configuration predict substantial contributions to both $v_c(R,z)$ and $a_z(R,z)$, resulting in a markedly better agreement with the data. 
We conclude that disk-like DM distributions are strongly favored over spherical halo models. 
Forthcoming \textit{Gaia} data releases will enable even more stringent tests of the geometry and distribution of the MW's DM component.
\end{abstract}

\section{Introduction}

The \emph{Gaia} mission, launched in 2013, is delivering an exceptionally precise and comprehensive 
three-dimensional map of more than a billion stars across the Milky Way (MW) and its immediate surroundings, 
while simultaneously providing accurate measurements of their kinematics 
\citep{Gaia_2016,Gaia_2018,Gaia_2021}. 
The third data release (DR3) of \emph{Gaia} has represented a major milestone, 
extending the catalog of stars with full six-dimensional phase-space information --- positions, 
parallaxes, proper motions, and line-of-sight velocities --- to over 30 million objects 
\citep{Katz_etal_2022}. 
The availability of such a vast and homogeneous dataset marks a transformative step in the study 
of Galactic dynamics, enabling precise determinations of the rotation curve and a wide range of 
kinematic and structural properties of the MW 
\citep{Antoja_etal_2021,Drimmel_etal_2022,Katz_etal_2022}. 

Analyses of several  stellar subsamples extracted from the most recent \emph{Gaia} data releases have revealed compelling evidence for a systematic decline in the MW's stellar rotation velocity \citep[e.g.][]{Eilers_etal_2019,Wang_etal_2023,Yongjun_etal_2023,Ou_etal_2024}.
Although this finding still awaits confirmation from forthcoming, more complete \emph{Gaia} data releases, it carries profound implications for models of the Galactic mass distribution.
The observed decrease in rotation velocity implies that the amount of dark matter (DM) in the MW is significantly smaller than predicted by models assuming a flat rotation curve \citep{SylosLabini_etal_2023,Ou_etal_2024}.
In particular, the declining rotation curve poses a direct challenge to the standard DM  halo paradigm --- represented by the Navarro-Frenk-White (NFW) profile \citep{Navarro_etal_1997} --- which was originally introduced to explain the approximately flat rotation curves observed in external spiral galaxies.

The evidence for a declining rotation curve in the Milky Way calls for a critical reassessment of the standard assumption that the Galactic DM halo is approximately spherical. If the DM contribution to the inner rotation curve is weaker than expected from such models, alternative geometries warrant consideration. A particularly compelling possibility is that a significant fraction of the DM is distributed in a flattened, disk-like structure that coexists with the baryonic disk \citep{SylosLabini_etal_2023}. In this configuration, the dark component contributes substantially to both the radial and vertical components of the gravitational field, thereby enhancing the off-plane rotation curve and the vertical acceleration. This dual influence may offer a viable alternative explanation for the observed off-plane kinematics and vertical structure of the Galaxy --- one that merits further investigation.

{    

Indeed, the MW offers a unique opportunity to investigate rotation curves beyond the Galactic plane.
In particular, the use of the \textit{Gaia} DR3 stellar sample within the spatial ranges
\( z \in (-2,2)\,\mathrm{kpc} \) and \( R \in (8.5,14)\,\mathrm{kpc} \)
to measure the kinematic properties of the local stellar population --- and to use these measurements to constrain dynamical mass models --- provides several important advantages.

\begin{enumerate}
\item \textit{High-precision kinematics:}
all three velocity components are measured with remarkable accuracy, with typical uncertainties of only a few \(\mathrm{km\,s^{-1}}\).

\item \textit{Simplified modelling:}
the relatively limited radial and vertical extent of the dataset allows for a substantially simplified description of both the stellar and DM density distributions. In particular, large-scale structural features of the Galactic disk, such as the \emph{warp} and \emph{flare}, which become important only at larger radii and heights, can be safely neglected.

\item \textit{Maximal dynamical gradients:}
the vertical variation of both the radial and vertical gravitational accelerations is strongest around
\( R \sim 10\,\mathrm{kpc} \), corresponding to approximately twice the characteristic scale length of the Galactic disk. This significantly enhances the sensitivity of dynamical diagnostics in this region.
\end{enumerate}
}

\noindent
{    

These properties allow us to improve the constraints on the geometry of the Galactic mass distribution recently obtained by
\cite{SylosLabini_2024,Lopez-Corredoira_2025}, who used the reconstructed dataset of \citet{Wang_etal_2023}, based on a statistical deconvolution of parallax errors through Lucy's inversion method (LIM).

In contrast, the present work directly fits the observational data for both the mid-plane and off-plane rotation curves, \(v_c(R,z)\), together with the vertical acceleration, \(a_z(R,z)\), derived from the \textit{Gaia} DR3 stellar sample with full six-dimensional phase-space information. As a result, the typical uncertainties on the rotation curve are reduced to
\(\sim 1\)--\(5\,\mathrm{km\,s^{-1}}\), compared to
\(\sim 8\)--\(20\,\mathrm{km\,s^{-1}}\) in the LIM-based analysis.

We consider two distinct mass models that share the same observed stellar and gaseous components, but differ in the geometry and total mass distribution of the DM component. The key difference between the models therefore lies in the assumed spatial distribution of DM, which has important consequences for the predicted gravitational field, particularly in the vertical direction.

The first model is the \emph{standard halo model}, in which the stellar and gaseous disks are embedded within a spherical DM halo \citep{Navarro_etal_1997}. The second is the \emph{dark matter disk model}, in which the DM is instead assumed to be distributed in a flattened, disk-like configuration sharing the same large-scale kinematic structure as the stellar and gaseous components \citep{Sancisi_1999,Hoekstra_etal_2001,Hessman+Ziebart_2011,Swaters_etal_2012,SylosLabini_etal_2024_Mass,SylosLabini_etal_2025}.

The paper is organized as follows. 
In Sect.~\ref{sect_jeans}, we briefly recall the Jeans equations that relate the observed kinematical 
quantities to the underlying gravitational potential, under the assumption of stationary equilibrium. 
Sect.~\ref{sect:data} describes the observational dataset employed in the analysis
and in Sect.~\ref{sect:massmod}, we present the mass models adopted for the various Galactic components. 
Sect.~\ref{sect:results} reports the results of the fitting procedure and the corresponding comparison 
between models and observations. 
Finally, Sect.~\ref{sect:concl} summarizes our main findings and discusses their broader implications.
}

%%%%%%%%%%%%%%%%%%%%%%%%%%%%%%%%%%%%%%%%%%%%%%%%%%%%%%%%%%%%%%%%%%%%%%%%%%%
%%%%%%%%%%%%%%%%%%%%%%%%%%%%%%%%%%%%%%%%%%%%%%%%%%%%%%%%%%%%%%%%%%%%%%%%%%%
%%%%%%%%%%%%%%%%%%%%%%%%%%%%%%%%%%%%%%%%%%%%%%%%%%%%%%%%%%%%%%%%%%%%%%%%%%%
%%%%%%%%%%%%%%%%%%%%%%%%%%%%%%%%%%%%%%%%%%%%%%%%%%%%%%%%%%%%%%%%%%%%%%%%%%%
%%%%%%%%%%%%%%%%%%%%%%%%%%%%%%%%%%%%%%%%%%%%%%%%%%%%%%%%%%%%%%%%%%%%%%%%%%%

\section{Jeans Equations in Cylindrical Coordinates}
\label{sect_jeans} 
{   

In a steady-state stellar system, the collisionless Boltzmann
equation reduces to the Jeans equations which relate velocity moments to the
gravitational potential \(\Phi \) \citep{BinneyTremaine2008}.
Before presenting their axisymmetric form
(Sects.~\ref{sec:radial_jeans}--\ref{sec:vertical_jeans}), we briefly recall
their derivation and clarify their interpretation in the context of Galactic
observations.

%%%%%%%%%%%%%%%%%%%%%%%%%%%%%%%%%%%%%%%%%%%%%%%%%%%%%%%%%%%%%%%%%%%%%%%%%%%

\subsection{Multi-component Jeans equations}
\label{sec:multicomp} 

We consider a collisionless system composed of particle species $s$ with mass
$m_s$. The microscopic phase-space density is
\begin{equation}
f_{\rm micro}(\mathbf{x},\mathbf{v},t)
= \sum_s \sum_{i\in s}
\delta^{(3)}(\mathbf{x}-\mathbf{x}_i)\,
\delta^{(3)}(\mathbf{v}-\mathbf{v}_i),
\end{equation}
which is replaced by coarse-grained distribution functions
$f_s(\mathbf{x},\mathbf{v},t)$ representing phase-space number densities.

Each component evolves according to the Vlasov equation,
\begin{equation}
\frac{\partial f_s}{\partial t}
+ \mathbf{v}\cdot\nabla f_s
- \nabla\Phi\cdot\nabla_{\mathbf{v}} f_s = 0,
\end{equation}
with the potential $\Phi$ sourced by the total mass density $\rho = \sum_s \rho_s=\sum_s n_s m_s$ (where $n_s$ the number density of the $s$-th species), and obeying  Poisson's equation
\begin{equation}
\nabla^2\Phi = 4\pi G \sum_s \rho_s.
\end{equation}
Taking velocity moments yields the Jeans equations for each component,

\begin{equation}
\frac{\partial}{\partial t}(\rho_s u_{s,i})
+ \frac{\partial}{\partial x_j}
\left(\rho_s u_{s,i}u_{s,j}+\rho_s\sigma^2_{s,ij}\right)
= -\rho_s \frac{\partial \Phi}{\partial x_i},
\end{equation}
and, summing over species, the combined equation
\begin{equation}
\frac{\partial}{\partial t}(\rho u_i)
+ \frac{\partial}{\partial x_j}
\left(\rho u_i u_j + \rho \sigma^2_{ij}\right)
= -\rho\,\frac{\partial \Phi}{\partial x_i}.
\label{eq:jeans_total}
\end{equation}

Dynamically distinct components must be defined by intrinsic quantities
conserved along orbits (e.g.\ particle mass). In general, observationally defined stellar
populations \textit{do not} satisfy this requirement and therefore do not correspond to
independent collisionless fluids.

%%%%%%%%%%%%%%%%%%%%%%%%%%%%%%%%%%%%%%%%%%%%%%%%%%%%%%%%%%%%%%%%%%%%%%%%%%%
\subsection{Connection to Galactic observations}
\label{sec:connection}
{
Dynamically independent collisionless components should strictly be defined through distribution functions conserved along particle orbits. Observational stellar samples (e.g., from \textit{Gaia}) generally do not satisfy this condition, since they are selected through observational cuts that are not orbit invariants. Their empirical distribution functions therefore cannot be interpreted as exact independent dynamical fluids.

In the present work, the observed stellar kinematics are instead used as tracer estimates of the coarse--grained velocity moments of the total gravitating system. Consequently, the density entering Eq.~(\ref{eq:jeans_total}) is interpreted as the total mass density sourcing the gravitational potential, consistently with the Poisson equation.

The resulting Jeans treatment should therefore be regarded as a phenomenological closure approximation rather than as an exact multi--component dynamical decomposition. In a general collisionless system, different components may possess different velocity dispersions, anisotropies, and asymmetric drifts. A fully multi--component treatment would thus require introducing additional phase--space properties for the dark matter component, which are not directly constrained observationally and would substantially increase the degeneracy of the problem.

By contrast, the combined Jeans equation provides a minimal effective closure in which observable stellar kinematics are used to probe the mean gravitational equilibrium of the system without introducing additional unconstrained dark matter kinematical degrees of freedom. The approach does not assume that stars and dark matter possess identical microscopic phase--space distributions, but rather that the observed stellar moments provide an effective large--scale description of the total dynamical state.

In this sense, the validity and limitations of the approximation must ultimately be assessed empirically through comparison with observations, rather than through an exact multi--component decomposition that is presently inaccessible observationally.
}

%%%%%%%%%%%%%%%%%%%%%%%%%%%%%%%%%%%%%%%%%%%%%%%%%%%%%%%%%%%%%%%%%%%%%%%%%%%

\subsection{The Radial Jeans Equation}
\label{sec:radial_jeans} 

Assuming stationarity, the radial Jeans equation reads
\begin{equation}
\label{eq:jeansR_full}
\frac{\partial (\rho \overline{v_R^2})}{\partial R}
+ \frac{\partial (\rho \overline{v_Rv_z})}{\partial z}
+ \rho\left(
\frac{\overline{v_R^2}-\overline{v_\phi^2}}{R}
+ \frac{\partial \Phi}{\partial R}
\right)
= 0.
\end{equation}
Defining $a_R=-\partial\Phi/\partial R$, a proxy of the circular velocity out of the equatorial ($z=0$) plane, 
$v_c^2(R,z)=-R a_R$ is
\begin{equation}
\label{eq:vc_jeans}
v_c^2(R,z)
= -R \left[
\frac{1}{\rho}\frac{\partial (\rho \overline{v_R^2})}{\partial R}
+ \frac{1}{\rho}\frac{\partial (\rho \overline{v_Rv_z})}{\partial z}
+ \frac{\overline{v_R^2}-\overline{v_\phi^2}}{R}
\right].
\end{equation}

This quantity feels mostly  the radial gravitational field. 
Only in the midplane ($z=0$) does it coincide with the velocity of circular orbits; at $|z|>0$,
vertical forces prevent strictly circular motion.

%%%%%%%%%%%%%%%%%%%%%%%%%%%%%%%%%%%%%%%%%%%%%%%%%%%%%%%%%%%%%%%%%%%%%%%%%%%

\subsection{The Vertical Jeans Equation}
\label{sec:vertical_jeans} 

The vertical Jeans equation is
\begin{equation}
\label{eq:jeansZ_full}
\frac{\partial (\rho \overline{v_Rv_z})}{\partial R}
+ \frac{\partial (\rho \overline{v_z^2})}{\partial z}
+ \rho\left(
\frac{\overline{v_Rv_z}}{R}
+ \frac{\partial \Phi}{\partial z}
\right)
= 0,
\end{equation}
which yields
\begin{equation}
\label{eq:aZ}
a_z = -\frac{\partial \Phi}{\partial z}
= \frac{1}{\rho}\frac{\partial (\rho \overline{v_Rv_z})}{\partial R}
+ \frac{1}{\rho}\frac{\partial (\rho \overline{v_z^2})}{\partial z}
+ \frac{\overline{v_Rv_z}}{R}.
\end{equation}

%%%%%%%%%%%%%%%%%%%%%%%%%%%%%%%%%%%%%%%%%%%%%%%%%%%%%%%%%%%%%%%%%%%%%%%%%%%
\subsection{Simplified form}

Neglecting the cross term $\overline{v_R v_z}$ in Eq. (\ref{eq:vc_jeans}) and Eq. (\ref{eq:aZ}) (e.g.\ \citealt{Eilers_etal_2019}), they reduce to 
\begin{align}
\label{eq:Jeans_eff}
v_c^2(R,z) &\simeq
- R \left[
\frac{1}{\rho}\frac{\partial (\rho \overline{v_R^2})}{\partial R}
+ \frac{\overline{v_R^2}-\overline{v_\phi^2}}{R}
\right], \\
a_z &\simeq
\frac{1}{\rho}\frac{\partial (\rho \overline{v_z^2})}{\partial z}.
\end{align}
These expressions will be adopted in the following analysis. Their use
implicitly relies on the mean-field closure discussed in
Sect.~\ref{sec:connection}, in which the velocity moments are interpreted as
tracer estimates of the total coarse-grained distribution function.
}

\section{Kinematic data analysis}
\label{sect:data} 
We first describe the properties of the stellar sample used in this analysis
(Sect.~\ref{sect:data_sample}).  We then determine the mean velocity components
as a function of radial distance (Sect.~\ref{sec:velocity_field}), followed by
the derivation of the rotation curve (Sect.~\ref{sec:rotation}).  We next turn
to the measurement of the vertical acceleration (Sect.~\ref{sec:vertical}).
Finally, we summarize and discuss our results in Sect.~\ref{sec:discussion_data}.

%%%%%%%%%%%%%%%%%%%%%%%%%%%%%%%%%%%%%%%%%%%%%%%%%%%%%%%%
\subsection{Description of the stellar sample}
\label{sect:data_sample}

The \emph{Gaia} mission has provided an unprecedentedly large and precise dataset for Galactic studies, 
delivering accurate measurements of stellar positions, parallaxes, and proper motions for more than 
one billion stars \citep{Gaia_2016,Gaia_2018,Gaia_2021}. 
The third data release (\emph{Gaia}~DR3) represents a major advancement, as it significantly expands the catalog 
of stars with line-of-sight velocity measurements --- providing full six-dimensional (6D) phase-space information 
for more than $3\times10^{7}$ objects \citep{Katz_etal_2022}. 
The availability of such complete kinematic data has opened a new era in the characterization of the MW's
rotation curve, velocity field, and other dynamical and structural properties 
\citep{Antoja_etal_2021,Drimmel_etal_2022,Katz_etal_2022}.

For the purposes of this study, we selected a subsample of \emph{Gaia}~DR3 stars with reliable astrometric and 
kinematic measurements, restricted to Galactocentric radii $R \leq 14$~kpc. 
This radial limit ensures that the relative uncertainty in distance remains below approximately $20\%$ 
\citep{Gaia_2021}, thereby guaranteeing a robust determination of spatial and velocity components. 
The resulting dataset contains roughly $1.6\times10^{6}$ stars with complete 6D phase-space information. 
To focus on the kinematics of the Galactic disk, we further restricted the sample to stars within a vertical height 
of $|z| < 2$~kpc from the midplane. 
We also excluded high-velocity outliers by applying cuts in the three cylindrical velocity components: 
$|v_R| < 50$~km\,s$^{-1}$, $|v_z| < 50$~km\,s$^{-1}$, and $v_\phi < 300$~km\,s$^{-1}$. 
These criteria remove stars with extreme kinematic properties, such as halo interlopers or stars affected 
by large measurement errors, thereby ensuring that the resulting sample is dominated by the dynamically 
cold population of the Galactic disk.

This selection provides a statistically significant representation of the local and intermediate regions of the disk, 
allowing us to probe the dynamical structure of the MW in the vicinity of the Sun with high precision.

\begin{figure*} 
\includegraphics[width=0.2450\textwidth]{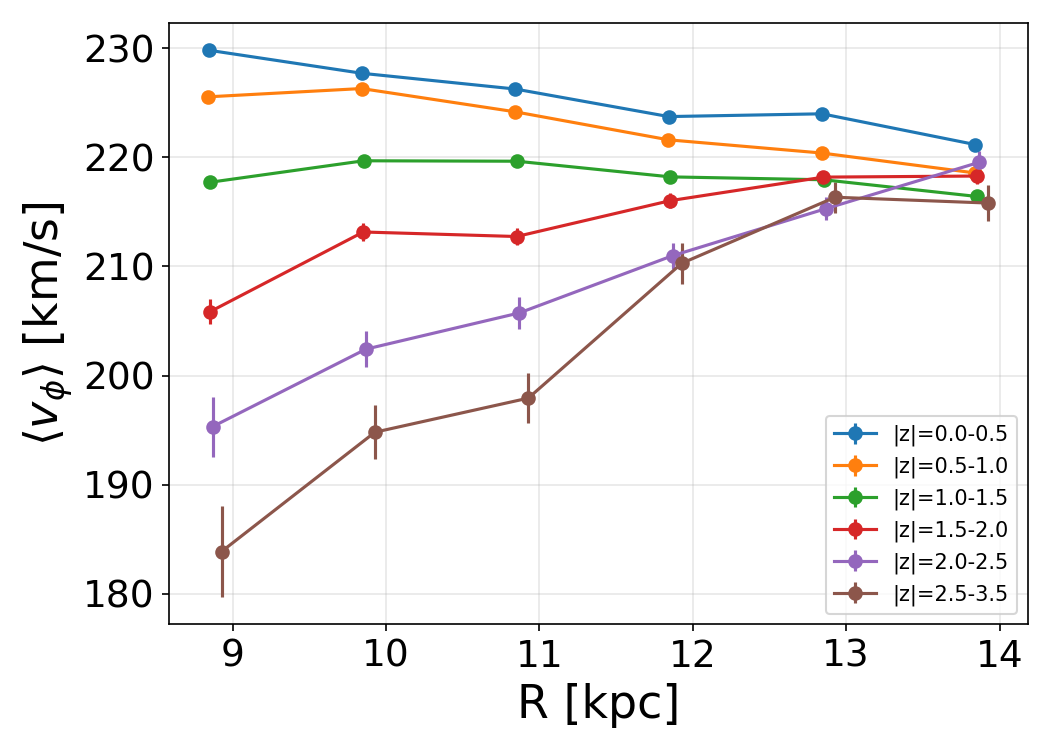}
\includegraphics[width=0.2450\textwidth]{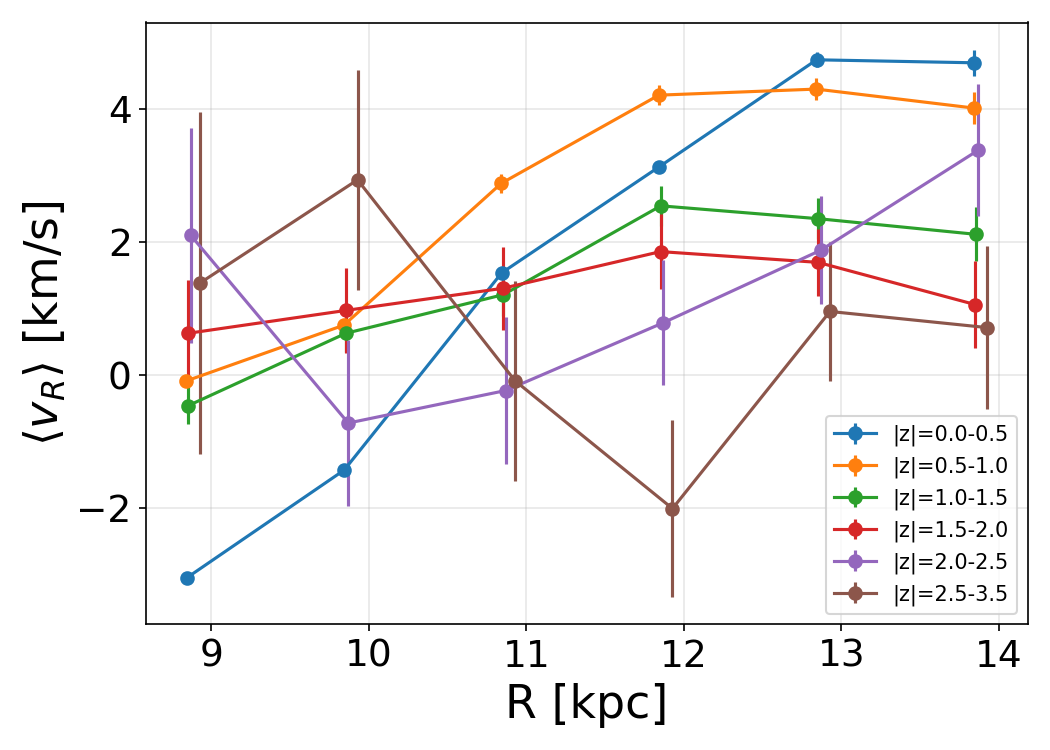}
\includegraphics[width=0.2450\textwidth]{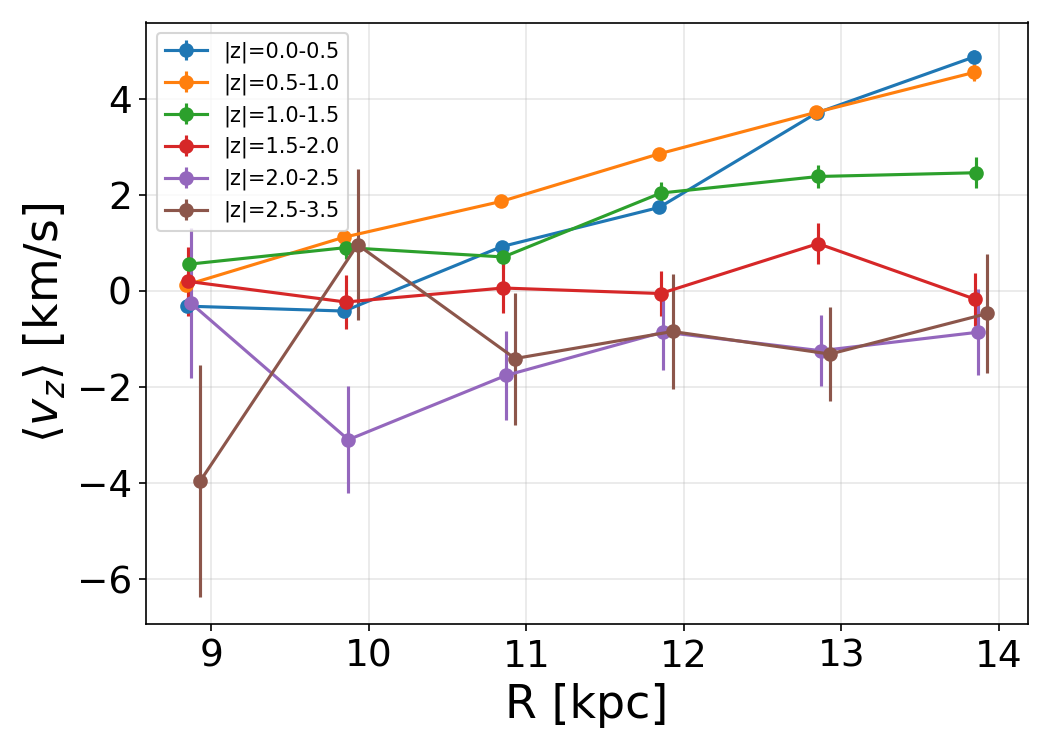}
\includegraphics[width=0.2450\textwidth]{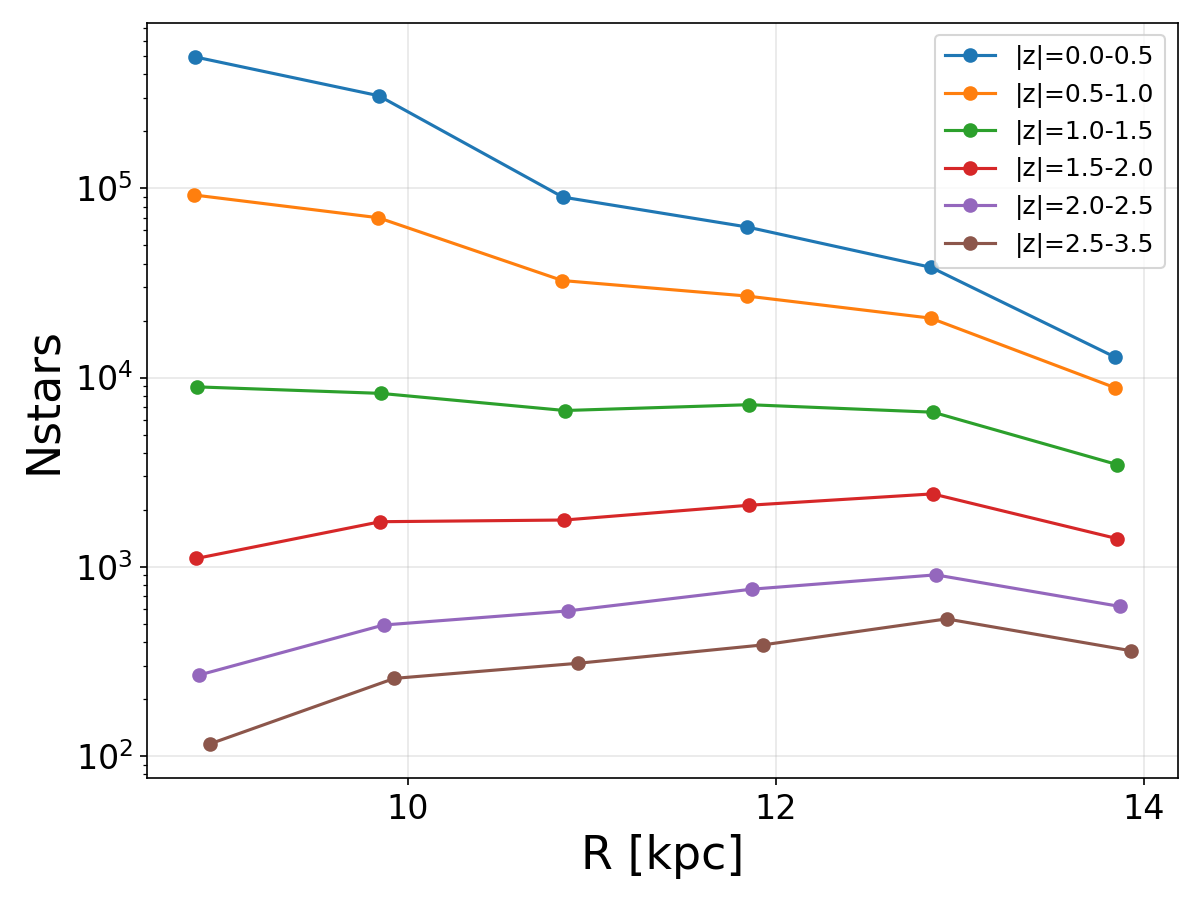}
\caption{   
Average radial profiles of the  three velocity components, $v_\phi$ (left panel), $v_r$  (middle left panel), and $v_z$ (middle right panel). 
	The right panel shows the number of stars per radial bin, one curve per $z$ slice.} 
\label{fig:radialprofiles} 
\end{figure*}

\begin{figure} 
    \centering
    \includegraphics[width=0.45\textwidth]{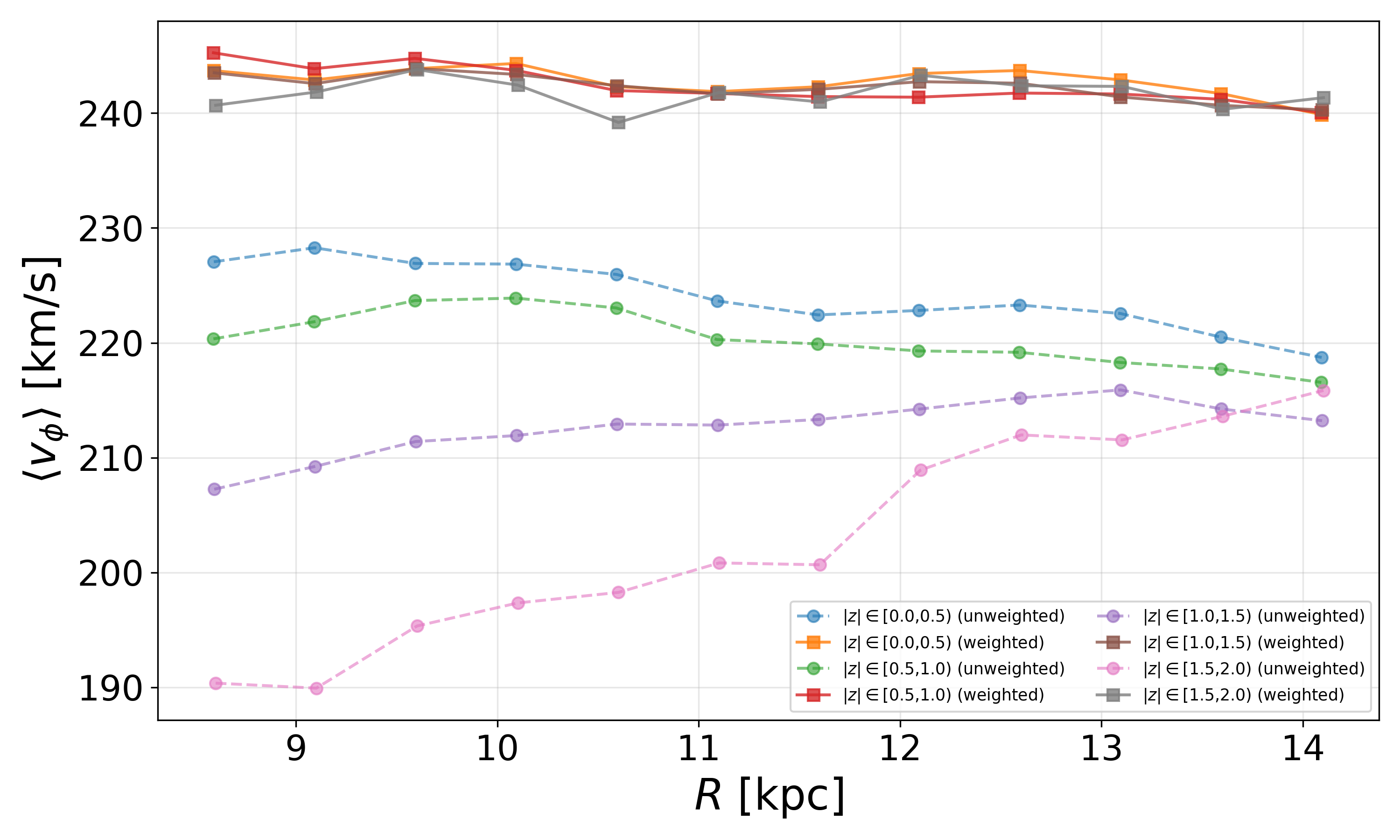}
    \caption{
        Comparison of weighted and unweighted mean azimuthal velocities $\langle v_\phi \rangle$ as a function of Galactocentric radius for four vertical slices in $|z|$.
        Stars with small errors ($\sigma(v_\phi) \sim 1$--2 km\,s$^{-1}$) tend to rotate faster ($v_\phi \sim 250$ km\,s$^{-1}$), 
        while stars with larger errors ($\sigma(v_\phi) \sim 5$--7 km\,s$^{-1}$) typically rotate more slowly ($v_\phi \sim 220$ km\,s$^{-1}$).
        Weighting by $1/\sigma^2$ therefore shifts the mean toward the kinematics of the low-error sub-population, introducing a systematic offset.
    }
    \label{fig:weighted-unweighted}
\end{figure}

%%%%%%%%%%%%%%%%%%%%%%%%%%%%%%%%%%%%%%%%%%%%%%%%%%%%%%%%

\subsection{Determination of the velocity field} 
\label{sec:velocity_field}

We divide the sample into six vertical slices, with the boundaries in $z$ and the corresponding number of stars reported in Tab.~\ref{tab1}. All slices are restricted to the radial range $(8.5, 14)$ kpc.
\begin{table}
\begin{center}
\begin{tabular}{| c | c | c | }
\hline
 $z_{\min}$  (kpc)&  $z_{\max}$  (kpc) & $N_{\rm s}$ \\ 
 \hline 
 0 & 0.5 & 1185943 \\  
 0.5 & 1.0 & 320786 \\  
 1.0 & 1.5 & 56724 \\  
 1.5 & 2.0 & 15797 \\  
 2.0 & 2.5 & 6141 \\  
 2.5 & 3.5 & 4256 \\   
\hline 
\end{tabular}
\end{center}
 \caption{Limits in the vertical height and number of stars in each of the 6 vertical slices}
 \label{tab1} 
\end{table}

The average value of each velocity component in a given bin can be computed as a weighted mean, using the reported uncertainties as weights:
\be
\bar{v} = \frac{\sum_i \frac{v_i}{\sigma_i^2}}{\sum_i \frac{1}{\sigma_i^2}} \;; 
\ee	
where the sum extends to all stars in a given spatial bin and the error on the mean can be obtained  from error propagation:
\be
\Delta \bar{v} = \left( \sum_i \frac{1}{\sigma_i^2} \right)^{-1/2}
\ee
where $\sigma_i$ is the measurement error.
When computing average velocities, one may adopt a  weighted mean, where each measurement $v_i$ is assigned a weight $w_i = 1/\sigma_i^2$, inversely proportional to the square of its quoted uncertainty:
this ensures that stars with smaller errors $\sigma_i$ contribute more significantly to the mean.
This weighting scheme is statistically justified if the uncertainties accurately reflect measurement quality ---  for example, if high signal-to-noise stars truly yield more reliable velocity estimates.

Alternatively, one may use the unweighted mean, which treats all stars equally regardless of their uncertainties:
\begin{equation}
\label{eq:unweighted}
\bar{v} = \frac{1}{N} \sum_i v_i,
\end{equation}
with an associated standard error of the mean given by:
\begin{equation}
\label{eq:unweighted2}
\Delta \bar{v} = \frac{\sqrt{\frac{1}{N - 1} \sum_i (v_i - \bar{v})^2}}{\sqrt{N}} \;.
\end{equation}
This approach is particularly appropriate when the aim is to characterize the overall kinematic properties of a stellar population, especially in situations where measurement uncertainties correlate with intrinsic stellar parameters (e.g., age, distance, or spectral type), which can introduce biases in the weighted estimate.

In \textit{Gaia} data, the weighted and unweighted means of $v_\phi$ often differ systematically --- not merely in their error bars, but in their central values.  
This discrepancy arises because the velocity uncertainties $\sigma(v_\phi)$ are not random: they correlate with stellar properties.  
Stars with smaller $\sigma(v_\phi)$ tend to be nearby, bright thin-disk stars that rotate faster, while fainter, more distant stars --- often belonging to older or dynamically hotter populations --- tend to have larger uncertainties and rotate more slowly.  
As a result, weighting by $1/\sigma^2$ biases the mean toward the kinematics of the low-error (young, cold) subset, while the unweighted mean better reflects the full mixture of stellar populations within a given radial and vertical bin.

For this reason, throughout our analysis we adopt the unweighted mean and its corresponding uncertainty as defined in Eqs.~\ref{eq:unweighted}--\ref{eq:unweighted2}, ensuring that our velocity profiles are representative of the entire stellar sample in each spatial bin, rather than just the best-measured --- but potentially biased --- subpopulation.

The upper panel of Fig.\ref{fig:radialprofiles} shows the average radial profiles of the  three velocity components, $v_R$, $v_\phi$, and $v_z$: in 
the bottom panel it is reported the number of stars per radial bin, one curve per $z$ slice. A clear systematic trend is visible: as $|z|$ increases, the mean value of $v_\phi$ progressively decreases everywhere but at for large values of the radial distance, i.e.  $R>13$ kpc.
On the other hand, the mean values of both $v_R$ and $v_z$ remain close to zero, with amplitude below $\sim 10$ km s$^{-1}$.
\begin{figure} 
 \centering
 \includegraphics[width=0.45\textwidth]{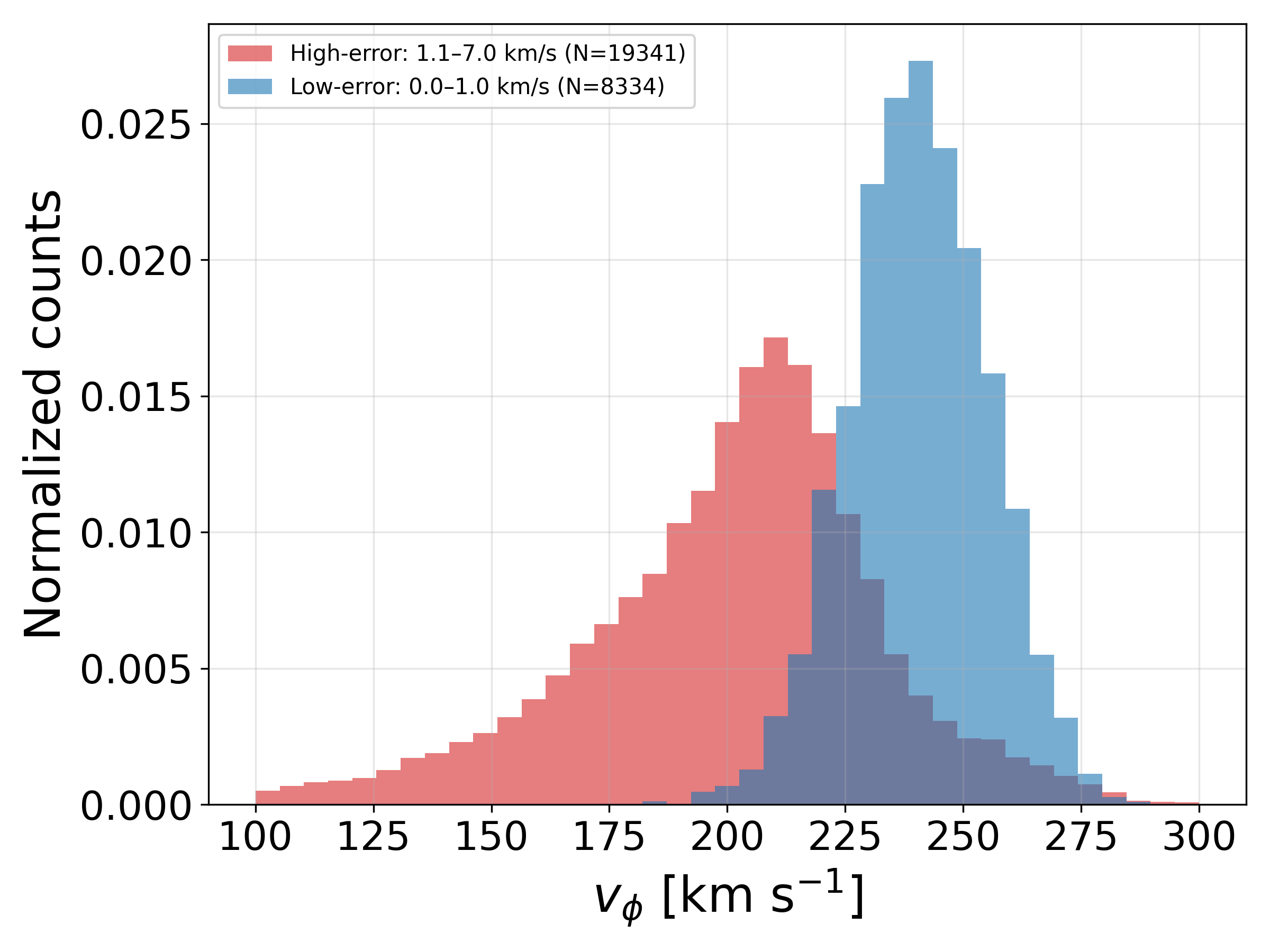}
 \caption{Comparison of the azimuthal velocity distributions $v_\phi$ for stars in the bin 
$9<R<11$~kpc and $1<|z|<2$~kpc, separated according to their velocity measurement uncertainties $\sigma(v_\phi)$. 
Stars with small uncertainties , i.e.  $0<\sigma(v_\phi)<1$ km\,s$^{-1}$ (blue distribution)  show a   peak is at $v_\phi \approx 250$~km\,s$^{-1}$
whereas stars with larger uncertainties , i.e.  $1.1<\sigma(v_\phi)<7$ km\,s$^{-1}$ (red  distribution)  show a   peak is at $v_\phi \approx 200$~km\,s$^{-1}$. 
}
\label{fig:vphi_histogram_low_vs_high_error}
\end{figure}

Figure~\ref{fig:weighted-unweighted} shows how this bias manifests across different vertical bins: stars with the smallest velocity uncertainties cluster around the highest rotation speeds, while those with larger uncertainties lag behind.  
This is a classic case of population-selection bias: in {\it Gaia} data, low-error stars tend to be bright, nearby, thin-disk members, while high-error stars are generally fainter, more distant, or belong to older, dynamically hotter disk populations exhibiting asymmetric drift.  
As a result, the weighted mean traces predominantly the cold, young disk, whereas the unweighted mean provides a more balanced representation of the full stellar mix.

Figure~\ref{fig:vphi_histogram_low_vs_high_error} shows the azimuthal velocity distributions $v_\phi$ for stars in the radial bin $9<$ kpc $R<11$~kpc and vertical bin $1<|z|<2$~kpc, separated according to their velocity measurement uncertainties $\sigma(v_\phi)$. 
Stars with small uncertainties ($0<\sigma(v_\phi)<1$~km\,s$^{-1}$; blue) form a distinct subpopulation, comprising only a minor fraction of the total sample. 
They exhibit a narrow velocity distribution centered at higher rotation speeds ($v_\phi \approx 250$~km\,s$^{-1}$), characteristic of younger, kinematically cold disk stars. 
Conversely, stars with larger uncertainties ($2<\sigma(v_\phi)<5$~km\,s$^{-1}$; red) rotate more slowly ($v_\phi \approx 200$~km\,s$^{-1}$) and show a broader distribution, typical of older, dynamically hotter populations. 
This behavior illustrates how inverse-variance weighting (i.e., weighting by $1/\sigma^2$) in the computation of mean velocities can bias the result toward the faster-rotating, low-error subset --- tracing mainly the young thin disk --- rather than the full underlying stellar population within the bin.

The observed systematic trend between $v_\phi$ and $\sigma(v_\phi)$ within the same $|z|$ bin can be interpreted in several ways:
\begin{enumerate}
    \item {Observational selection within the bin.}  
    Even within a fixed radial and vertical bin, {\it Gaia} uncertainties correlate with magnitude and crowding:  
    (i) Bright main-sequence stars have smaller $\sigma(v_\phi)$, are typically younger and dynamically colder, and rotate closer to the circular speed;  
    (ii) Fainter dwarfs or evolved giants have larger $\sigma(v_\phi)$, and are more likely to be older disk stars with higher velocity dispersion and asymmetric drift.  
    Thus, the internal distribution of $\sigma(v_\phi)$ acts as a proxy for stellar population age and kinematics.

    \item {Asymmetric drift within the thin disk.}  
    Even among thin-disk stars, older or dynamically hotter populations rotate more slowly due to asymmetric drift {    (see Eq.\ref{eq:Jeans_eff})}:
    \begin{equation}
        \overline{v_\phi} \approx v_c - \frac{\sigma_R^2}{2 v_c} 
        \left( 1 + \frac{\partial \ln \rho}{\partial \ln R} + \frac{\partial \ln \sigma_R^2}{\partial \ln R} \right),
    \end{equation}
    where $v_c$ is the circular velocity and $\sigma_R$ the radial velocity dispersion.  
    If older stars have larger measurement errors (e.g., due to being intrinsically fainter), this leads to exactly the observed trend:  
    stars with $\sigma(v_\phi) \sim 1$--2 km\,s$^{-1}$ cluster around $v_\phi \sim 250$ km\,s$^{-1}$, while stars with $\sigma(v_\phi) \sim 5$--7 km\,s$^{-1}$ exhibit $v_\phi \sim 220$ km\,s$^{-1}$.

    \item {Hidden metallicity and age biases.}  
    {\it Gaia} radial velocity uncertainties correlate with spectral type and luminosity class \citep[e.g.,][]{Yu+Liu_2018,Katz_etal_2019,Elsanhoury_etal_2023}.  
    F-G dwarfs (typically young and metal-rich) tend to yield smaller $\sigma(v_\phi)$ and rotate faster,  
    while K-M dwarfs (older and cooler) yield larger uncertainties and rotate more slowly.  
    Thus, the observed correlation between $v_\phi$ and $\sigma(v_\phi)$ indirectly reflects the underlying age-metallicity-kinematics relation.
\end{enumerate}

In conclusion, the offset between the weighted and unweighted means is not a
numerical artifact but the natural kinematic signature of stellar population
mixing.  Even in regions dominated by the thin disk, stars of different ages
and velocity dispersions rotate at systematically different speeds: younger,
dynamically colder stars rotate near $v_\phi \simeq 250$~km\,s$^{-1}$, whereas
older, dynamically hotter stars lag behind at $v_\phi \simeq 220$~km\,s$^{-1}$.
The observed correlation between $v_\phi$ and $\sigma(v_\phi)$ therefore
encodes genuine information on disk heating and the secular evolution of the
stellar population, rather than being merely a by-product of thick-disk or
halo contamination.

For this reason, throughout our analysis we adopt the unweighted mean and its
corresponding uncertainty (Eqs.~\ref{eq:unweighted}--\ref{eq:unweighted2}) as
our estimator of the stellar kinematic profiles.  This choice is fully
consistent with the mean-field framework underlying the dynamical mass models,
which rely on the \emph{total} coarse-grained distribution function and the
corresponding Jeans equation discussed in Sect.~\ref{sec:multicomp}
(see Eq.~\ref{eq:jeans_total}).

%%%%%%%%%%%%%%%%%%%%%%%%%%%%%%%%%%%%%%%%%%%%%%%%%%%%%%%%
%%%%%%%%%%%%%%%%%%%%%%%%%%%%%%%%%%%%%%%%%%%%%%%%%%%%%%%%

\subsection{Rotation Curve Estimate with Asymmetric Drift Correction}
\label{sec:rotation}

Let us now discuss the derivation of the rotation curve estimated through the asymmetric drift correction.  
In principle, to solve the Jeans equations (Eqs.~\ref{eq:vc_jeans}--\ref{eq:jeansZ_full}), one should know the full spatial dependence of the mass density, $\rho(R,z)$.  
In a realistic scenario, $\rho(R,z)$ is the sum of multiple components (stellar disks, gas, bulge, and DM), and its derivatives therefore depend on several structural parameters.  
To simplify the problem, we estimate the circular velocity $v_c(R,z)$ and the vertical acceleration $a_z(R,z)$ by solving the axisymmetric Jeans equations under the assumption that the total mass distribution follows a double-exponential disk profile of the form:
\be
\label{eq:double_exp} 
\rho(R,z) = \rho_0 \exp\left(-\frac{R}{R_{\rm d}}  \right) \exp\left (-\frac{|z|}{z_{\rm d}}\right) \;,
\ee
{    
where $R_d$ and $z_d$ are respectively the radial and vertical characteristic length scale of the disk. 
}
This approximation accurately describes the mass distribution within the
relatively limited range of galactocentric distances and heights relevant for
this study.

The Jeans equation (see Eq.\ref{eq:Jeans_eff}) then gives
\be
\label{eq:rotation_velocity} 
v_c^2(R,z) = \overline{v_\phi}^2 - \sigma_\phi^2 -
\sigma_R^2 \left( 1 + \frac{\partial \ln (\rho \sigma_R^2)}{\partial \ln R } + \frac{\partial \ln \rho}{\partial \ln R}\right),
\ee
where, in general we have  \( \overline{v^2} = \overline{v}^2 -\sigma^2 \) and where 
$ \overline{v_\phi}$, $\sigma_\phi$, and $\sigma_R$ are measured directly from the data.  
The density gradient follows from the exponential form, 
\[
\frac{\partial \ln \rho}{\partial \ln R} = -\frac{R}{R_{\rm d}},
\]
while 
\[ \frac{\partial \ln \sigma_R^2 }{ \partial \ln R} \] 
 is evaluated numerically after smoothing.  

The strategy of the estimation is:   
(i) Bin the data radially ($\Delta R = 0.5$ kpc).  
(ii) In each bin, compute $ \overline{v_\phi}$, $\sigma_\phi^2$, and $\sigma_R^2$.  
(iii) Insert these into the Jeans equation to obtain $v_c(R)$ for each $|z|$ slice.  

For the error estimation we adopt unweighted means and dispersions to reflect population averages. Formal Gaussian propagation yields an approximate uncertainty,
\[
\Delta v_c \approx \frac{1}{2 v_c} \sqrt{ (2 \overline{v_\phi} \Delta \langle v_\phi \rangle)^2 
+ (2\sigma_\phi \Delta \sigma_\phi)^2 
+ (2\sigma_R \Delta \sigma_R)^2 },
\]
but this underestimates the true error. A more robust approach is bootstrap resampling: in each radial bin we resample the stars (e.g. 200 times), recompute $v_c$, and adopt the standard deviation of these realizations as the error bar. This naturally accounts for covariances, small-$N$ noise, and fluctuations in the gradient term.  
Thus, for each $|z|$ slice we obtain $v_c(R)$ with bootstrap-based uncertainties, and we compare all six slices to test for vertical trends in the rotation curve. 

{   
The very small error bars obtained from bootstrap resampling are a tell-tale sign that the estimate is dominated by pure resampling noise, scaling as $1/\sqrt{N}$. With $N \sim 10^5$-$10^6$ stars, the statistical error therefore becomes negligible.
The true astrophysical uncertainty, however, is much larger, for several reasons:
(i) Systematics dominate: the Jeans equation involves density gradients and velocity-dispersion terms, and even small variations in these quantities can change $v_c$ by several km s$^{-1}$.
(ii) Bootstrap ignores correlated structures: if the sample contains substructure or non-Gaussian tails, resampling from the same dataset does not properly capture the associated uncertainty.
(iii) Bin-to-bin covariance: adjacent radial bins are not independent, so the variance across bins often provides a more realistic estimate of the uncertainty.
}

To avoid underestimating uncertainties, one should introduce an ``error floor'' that reflects astrophysical and systematic effects, not just Poisson noise. Possible approaches include:
(i) Estimating the scatter between neighboring bins and adopting 
    \[
    \Delta v_c(R_i,z) \approx \mathrm{RMS} \big( v_c(R_{i+1},z) - v_c(R_i,z) \big)
    \]
    as the error bar.  
(ii) Performing jackknife resampling by \emph{sky region} (dividing the stars spatially, rather than randomly), which typically yields larger, more realistic uncertainties.  
(iii) Inflating the bootstrap error by adding in quadrature a term proportional to $\sigma/\sqrt{N_{\mathrm{eff}}}$, where $N_{\mathrm{eff}}$ is an effective number of independent tracers (e.g. a few thousand, rather than the full $10^6$).  
(iv)Reporting both: the formal bootstrap errors (tiny) and the empirical scatter computed (from bin-to-bin scatter), and explicitly explaining the difference.

\begin{figure} 
\includegraphics[width=0.45\textwidth]{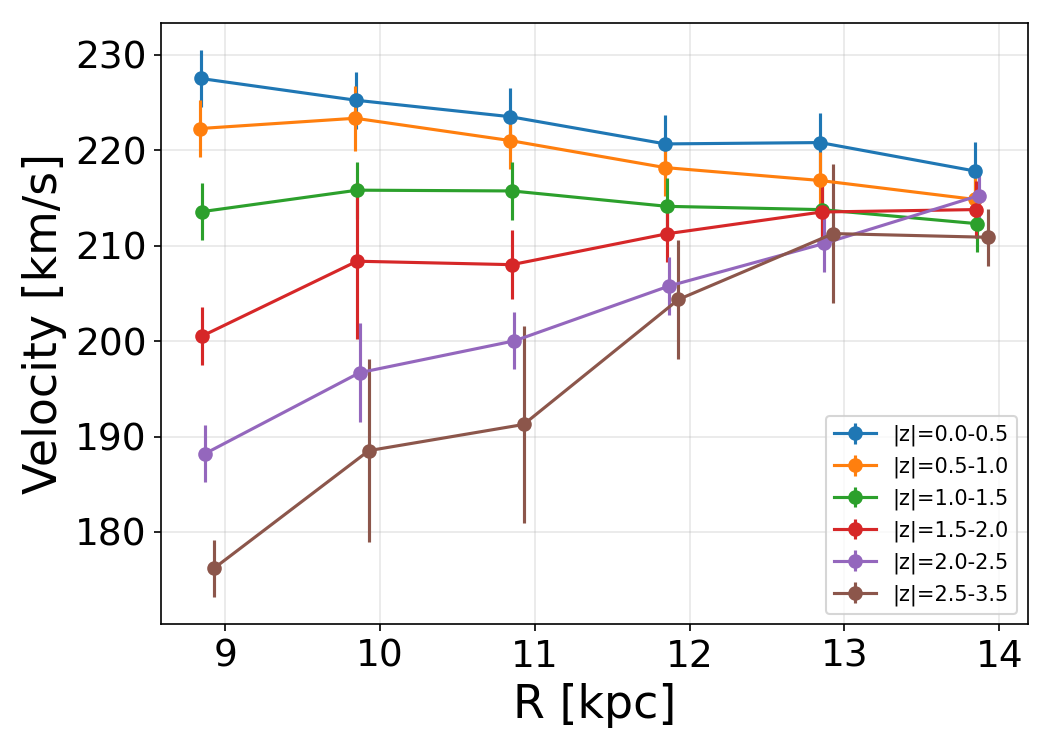}
	\caption{The rotation curves with error bars  in the different 6 vertical slices for $R_{\text{d}}=4.5$ kpc. } 
\label{fig:rotation_curves} 
\end{figure}

In practice, this procedure ensures that quoted uncertainties on $v_c(R)$ reflect real astrophysical limitations rather than the artificially small statistical noise of very large samples. 
With Gaia-scale datasets, the formal errors become negligible, yet the true uncertainties are dominated by systematics: non-Gaussian velocity distributions, correlated stars, noisy dispersion gradients, and deviations from the simplifying assumptions. 
Even after including the empirical ``bin-to-bin scatter,'' one still obtains unrealistically small values ($\sim 0.1$-1 km s$^{-1}$).
Following common practice, we impose a minimum uncertainty ("error floor") on $v_c$  to reflect astrophysical and modeling systematics that dominate over formal statistical errors in large samples (e.g., \cite{Hogg_etal_2010,Eilers_etal_2019}) 
\[
\Delta v_c = \max \left( \Delta v_c^{\mathrm{stat}}, \; \Delta v_c^{\mathrm{min}} \right),
\]
where $\Delta v_c^{\mathrm{stat}}$ is the statistical error (bootstrap or empirical) and $\Delta v_c^{\mathrm{min}}$ a fixed minimum, typically $\sim 3$ km s$^{-1}$. 
This acknowledges that systematics dominate and that the achievable precision cannot realistically surpass a few km s$^{-1}$, regardless of sample size. 
Notably, this uncertainty is of the same order as the asymmetric drift correction itself, highlighting that the dominant limitation is astrophysical rather than statistical.  
The resulting profiles are shown in Fig.~\ref{fig:rotation_curves}. Note that, in what follows, the fitting is restricted to the first four $|z|$ slices, since for $|z|>2$ kpc the stellar counts drop rapidly and the resulting profiles become dominated by statistical fluctuations.

%%%%%%%%%%%%%%%%%%%%%%%%%%%%%%%%%%%%%%%%%%%%%%%%%%%%%%%%
%%%%%%%%%%%%%%%%%%%%%%%%%%%%%%%%%%%%%%%%%%%%%%%%%%%%%%%%
%%%%%%%%%%%%%%%%%%%%%%%%%%%%%%%%%%%%%%%%%%%%%%%%%%%%%%%%
%%%%%%%%%%%%%%%%%%%%%%%%%%%%%%%%%%%%%%%%%%%%%%%%%%%%%%%%
%%%%%%%%%%%%%%%%%%%%%%%%%%%%%%%%%%%%%%%%%%%%%%%%%%%%%%%%

\subsection{Computation of the Vertical Acceleration and Its Uncertainty}
\label{sec:vertical}

We estimate the vertical acceleration $a_z(R,z)$ of stars in the Galaxy at fixed Galactocentric radii $R$ by applying the vertical Jeans equation to the observed vertical velocity dispersion of a tracer population.
If the vertical mass density of the disk follows an exponential profile (see Eq.~\ref{eq:double_exp}), the vertical Jeans equation  (see Eq.\ref{eq:Jeans_eff}) simplifies to:
\begin{align}
  a_z(R,z)
  &= \frac{1}{\rho(z)}\frac{\partial}{\partial z}\left[\rho(z)\,\sigma_z^2(R,z)\right] \\
  \nonumber 
  &= \frac{\partial \sigma_z^2(R,z)}{\partial z}
    -\frac{\mathrm{sign}(z)}{z_{\text{d}}}\,\sigma_z^2(R,z)\,,
  \label{eq:azKG2}
\end{align}
where $\sigma_z^2(R,z)$ is the vertical velocity dispersion squared, computed in bins of vertical height $z$ at fixed radius $R$, and $z_{\text{d}}$ is the vertical scale height.

The vertical derivative $\partial \sigma_z^2/\partial z$ is estimated numerically using a centered finite-difference method over the bin width $\Delta z$.
In each $(R,z)$ bin, the vertical velocity dispersion is computed as the sample variance:
\begin{equation}
  \sigma_z^2(R,z) = \frac{1}{N-1} \sum_{i=1}^N \left[v_{z,i}(R,z) - \overline{v_z(R,z)} \right]^2,
\end{equation}
where $N$ is the number of stars in the bin. Bins with fewer than a minimum threshold of stars are excluded from the analysis to ensure statistical reliability.

The associated uncertainty on the variance is estimated via the standard sampling error:
\begin{equation}
  \delta\sigma_z^2(R,z) = \sigma_z^2(R,z) \sqrt{\frac{2}{N - 1}}\,.
\end{equation}

The total propagated uncertainty on the vertical acceleration is:
\bea
\label{eq:az_full_error}
&&(\Delta a_z)^2 =
\left[
\Delta \left( \frac{\partial \sigma_z^2}{\partial z} \right)\right]^2
\\ \nonumber && 
+ \left( \frac{\mathrm{sign}(z)}{z_{\text{d}}} \, \Delta(\sigma_z^2) \right)^2
+ \left( \frac{\mathrm{sign}(z)\,\sigma_z^2}{z_{\text{d}}^2} \, \Delta z_{\text{d}} \right)^2.
\eea

In relative terms, if the uncertainties are small and mutually independent, 
the fractional uncertainty on the vertical acceleration can be approximated as
\begin{equation}
\label{eq:az_full_error_rel}
\frac{\Delta a_z}{|a_z|} \;\simeq\;
\sqrt{
\left( \frac{\Delta\!\left( \partial \sigma_z^2 / \partial z \right)}{a_z} \right)^{\!2}
+ \left( \frac{\Delta(\sigma_z^2)}{\sigma_z^2} \right)^{\!2}
+ \left( \frac{\Delta z_{\mathrm{d}}}{z_{\mathrm{d}}} \right)^{\!2}
}\,.
\end{equation}
The first term in Eq.~\ref{eq:az_full_error_rel} represents the uncertainty propagated from the 
vertical gradient of the velocity dispersion $\sigma_z^2$, 
the second term accounts for the uncertainty in the dispersion amplitude itself, 
and the third term quantifies the effect of uncertainty in the assumed vertical scale height $z_{\mathrm{d}}$. 
The latter typically corresponds to a systematic component that tends to dominate the total 
error budget on $a_z$.

The resulting profiles $a_z(R,z) \pm \delta a_z(R,z)$ provide a quantitative estimate of the vertical gravitational acceleration and its uncertainty across different heights and radii in the Galactic disk.

\begin{figure} 
\centering
\includegraphics[width=0.46\textwidth]{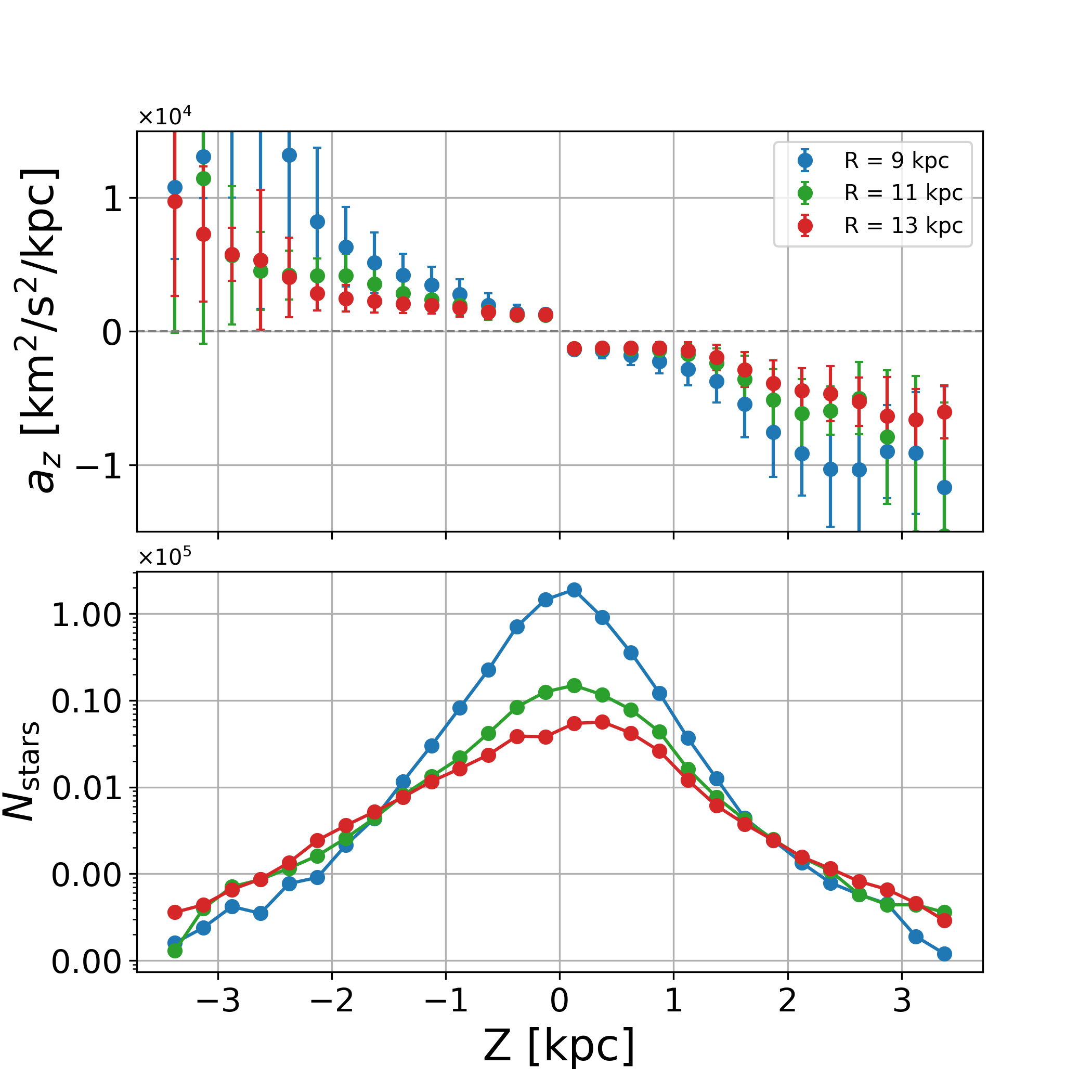}
\includegraphics[width=0.42\textwidth]{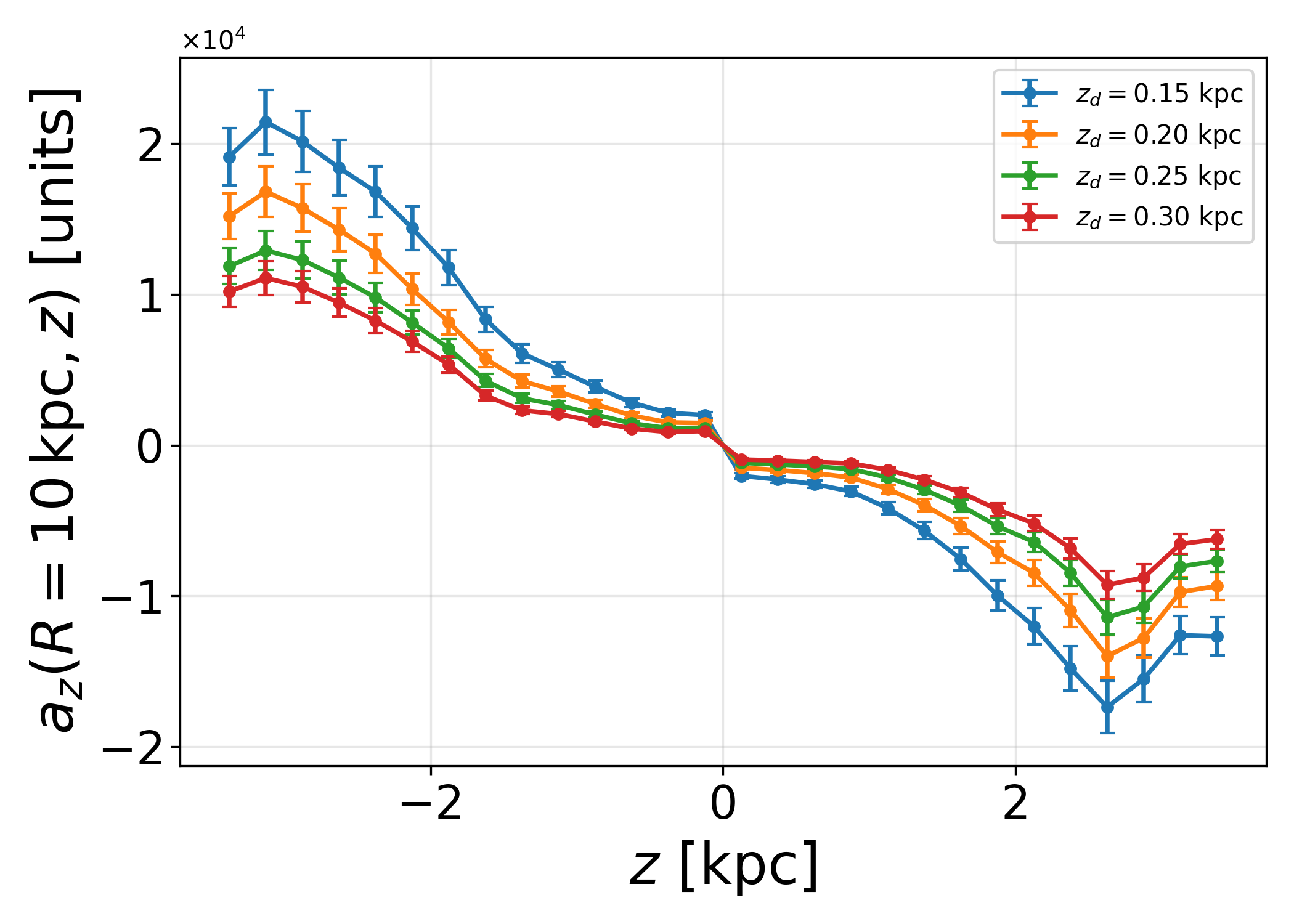}	
\caption{Upper panel: vertical acceleration and number of stars contributing to the average in function of $z$ for different $R$:  vs $z$ for  $R=9, 11, 13$ kpc and $z_{\text{d}}=0.25$ kpc. 
	Bottom panel: vertical acceleration for $R=10$ and different  values of $z_{\text{d}}$.
	} 
\label{fig:vertical_acceleration_1} 
\end{figure}

\subsection{Discussion} 
\label{sec:discussion_data}
Under the assumption that the stellar disk follows a double-exponential density profile, the rotation curve $v_c(R,z)$ depends primarily on the azimuthal velocity $\overline{v_\phi}$, with corrections that depend on the radial scale length $R_{\text{d}}$ (see Eq.~\ref{eq:rotation_velocity}). These corrections are typically subdominant: for instance, changing $R_{\text{d}}$ by a factor of two alters $v_c(R,z)$  by only  $\sim 3~\mathrm{km\,s^{-1}}$.

In contrast, the vertical acceleration $a_z(R,z)$ is particularly sensitive to the assumed vertical scale height $z_{\text{d}}$. Since $a_z(R,z) \propto 1/z_{\text{d}}$, any uncertainty in $z_{\text{d}}$ propagates linearly into the estimated acceleration. Specifically, if $z_{\text{d}} \rightarrow \alpha\, z_{\text{d}}$, then $a_z(R,z) \rightarrow a_z(R,z)/\alpha$. Therefore, a 50\% uncertainty in $z_{\text{d}}$ translates into a corresponding 50\% uncertainty in $a_z(R,z)$.

This sensitivity poses a significant challenge: the vertical scale heights of the thin and thick disks differ by nearly an order of magnitude, and their relative contributions vary with both Galactocentric radius and vertical height. Accurately modeling the vertical gravitational acceleration therefore requires precise constraints on the vertical structure of each disk component. Moreover, in the case of DM disk models, the vertical scale height can, in principle, be treated as a free parameter, to be determined by best fitting the observed vertical acceleration profiles.

{   

We note that the dataset used in this work is fundamentally different from that employed by \cite{SylosLabini_2024,Lopez-Corredoira_2025}. 
In those analyses, the kinematical information was not derived directly from individual \textit{Gaia} DR3 stars, but from the reconstructed dataset of \citet{Wang_etal_2023}. 
This dataset was constructed by applying a statistical deconvolution of parallax errors using Lucy's inversion method (LIM), producing binned kinematical maps in $(R,z)$ cells. 
Each cell contains many stars, and the velocity moments are estimated as averages within the cell. 
This approach is specifically designed to probe large Galactocentric distances ($R > 18\,\mathrm{kpc}$), where individual parallax errors are large ($\gtrsim 20\%$), by reducing statistical noise through averaging.

In contrast, the present work uses a directly selected sample of \textit{Gaia} DR3 stars with controlled quality cuts, without applying any statistical deconvolution or binning-based reconstruction. 
The kinematical quantities are therefore measured directly from the observed stellar sample, rather than inferred through an inversion procedure.

This difference has a clear quantitative impact on the uncertainties. 
In the present analysis, the typical uncertainties on the rotation curve are of order $\sim 1$--$5\,\mathrm{km\,s^{-1}}$, whereas in the LIM-based reconstruction they are significantly larger, $\sim 8$--$20\,\mathrm{km\,s^{-1}}$. 
This improvement by a factor of $\sim 3$--$5$ in kinematical precision directly translates into stronger constraints on the mass models, as discussed in the manuscript.

In addition, the present sample is more homogeneous in terms of selection, as it avoids the mixing of stars with widely different distance uncertainties inherent in the LIM reconstruction, and does not rely on cell-based averaging. 
This allows for a more direct and controlled comparison between the observed kinematics and the Jeans-based modeling.

}

%%%%%%%%%%%%%%%%%%%%%%%%%%%%%%%%%%%%%%%%%%%%%%%%%%%%%%%%%%%%
%%%%%%%%%%%%%%%%%%%%%%%%%%%%%%%%%%%%%%%%%%%%%%%%%%%%%%%%%%%%
%%%%%%%%%%%%%%%%%%%%%%%%%%%%%%%%%%%%%%%%%%%%%%%%%%%%%%%%%%%%
%%%%%%%%%%%%%%%%%%%%%%%%%%%%%%%%%%%%%%%%%%%%%%%%%%%%%%%%%%%%
%%%%%%%%%%%%%%%%%%%%%%%%%%%%%%%%%%%%%%%%%%%%%%%%%%%%%%%%%%%%
%%%%%%%%%%%%%%%%%%%%%%%%%%%%%%%%%%%%%%%%%%%%%%%%%%%%%%%%%%%%

\section{Mass models}
\label{sect:massmod}

The first model we consider is the \emph{standard halo model}, in which the stellar and gaseous disks are embedded in a spherical  DM halo. 
In this case, the DM particles are assumed to be in dynamical equilibrium under their own self-gravity, resulting in a quasi-isotropic velocity distribution.
We adopt the standard NFW  profile \citep{Navarro_etal_1997}, which has been widely used in the literature. 
More recently, cored halo profiles have also been employed to model galaxy rotation curves. 
For example, \citet{Ou_etal_2024} fit the Milky Way rotation curve using both a generalized NFW profile and an Einasto profile 
\citep[see also][]{Merritt_etal_2006}, each introducing an additional shape parameter beyond the standard NFW model. 
These more flexible models are particularly suited to reproducing a declining rotation curve, as observed in the outer Galaxy 
\citep{Eilers_etal_2019,Wang_etal_2023,Ou_etal_2024}.
However, since our analysis focuses on an intermediate radial range --- where the rotation curve is neither declining nor showing a significant central core --- the standard NFW profile is sufficient to capture the essential features of the DM halo, without introducing additional free parameters.

The second is the \emph{dark matter disk model}, in which the DM is hypothesized to be distributed in a flattened, disk-like configuration 
that shares the same kinematic structure as the stellar and gaseous components. 
The key features of this model are, first, the hypothesis that the DM spatial distribution follows that of the 
H\,\textsc{i} gas --- whose density declines more slowly with radius than that of the stellar component --- and, second, 
that the gravitational potential of a disk is stronger than that of a spherical mass distribution. 
As a result, a disk-like configuration can reproduce the same contribution to the circular velocity with a smaller 
total mass.

The key difference between these two models lies in the geometry of the DM distribution and, 
consequently, in the total disk mass and in the behavior of the gravitational potential gradient along the vertical direction. 
In the {halo model}, where DM is assumed to be spherically distributed, the vertical properties of the gravitational field 
are determined mainly by the baryonic components --- namely, the stellar and gaseous disks  --- whose spatial distributions are constrained by observations. 
In contrast, in the {disk model}, the DM itself is distributed within the Galactic disk and therefore contributes directly to shaping 
the vertical structure of the gravitational potential. 
As we show below, this makes the DM disk model particularly effective in reproducing the observed off-plane rotation curves 
and vertical accelerations.

The roots of the idea that the dynamically relevant part of DM can be located on the disk trace back to the observed correlation between DM and the distribution of neutral hydrogen (H\,\textsc{i}) gas, 
first noted by \citet{Bosma_1978,Bosma_1981}. 
In a sample of nearby disk galaxies, Bosma found that the ratio between the total disk surface density 
(derived from rotation curves) and the gas surface density (measured from H\,\textsc{i} observations) 
remains approximately constant beyond about one-third of the optical radius. 
This correlation implies that, at large radii, rotation curves can be viewed as rescaled versions of those obtained 
from the H\,\textsc{i} gas distribution. Such a relatively simple model works well in the case of a number of external galaxies as found by 
\cite{Sancisi_1999,Hoekstra_etal_2001,Hessman+Ziebart_2011,Swaters_etal_2012,SylosLabini_etal_2024_Mass,SylosLabini_etal_2025}. 

In this framework the total disk surface density is dominated by the stellar component 
at small radii, while at large radii it is dominated by a mass component that scales with the gas surface density. 
Since the gas mass constitutes only a small fraction of the total galactic mass, the Bosma effect suggests that 
the gas surface density acts as a proxy for the DM distribution. 
This, in turn, implies that DM follows the spatial distribution of the gas and is confined to a disk, 
even though it is not directly observable. 

{   
In both mass models, the total gravitational potential of the Galaxy is modeled as the sum of stellar plus gas (S)  and DM contributions,
}
\[
\Phi(R,z) \;=\; \Phi_{\rm S}(R,z) \;+\; \Phi_{\rm DM}(R,z).
\]
The stellar and gas potential is composed of several components constrained by observations,
\[
\Phi_{\rm S}(R,z)\;=\; 
\Phi_{\rm tnd} \;+\; 
\Phi_{\rm tkd} \;+\; \\
\Phi_{\rm  H \textsc{i}} \;+\; 
\Phi_{\rm b},
\]
where $\Phi_{\rm tnd}(R,z)$ corresponds to the thin disk, $\Phi_{\rm tkd}(R,z)$ to the thick disk, $\Phi_{\rm H \textsc{i}}(R,z)$ to the neutral hydrogen disk, and $\Phi_{\rm b}(R,z)$ to the bulge.  

For the DM distribution we consider two alternative models. The first is the standard NFW halo,  
$
\Phi_{\rm DM}(R,z) \;=\; \Phi_{\rm h}(R,z),
$
while the second is a DM disk (DMD),  
$
\Phi_{\rm DM}(R,z) \;=\; \Phi_{\rm m}(R,z).
$

Since the equations are linear in the potential, the contributions to the squared circular velocity are additive:
\bea
v_c^2(R,z) 
&=& v_{c,{\rm S}}^{\,2}(R,z) \;+\; v_{c,{\rm DM}}^{\,2}(R,z).
\eea

For the stellar components specifically, one has
\bea
v_{c,{\rm S}}^{\,2}(R,z) 
&=& 
v_{c,{\rm tnd}}^{\,2}
+ v_{c,{\rm tkd}}^{\,2}
%\\ \nonumber 
%&+& 
+ v_{c,{\rm H \textsc{i}}}^{\,2} 
+ v_{c,{\rm b}}^{\,2}.
\eea

We begin by discussing the stellar and gaseous components of the Galaxy, whose properties are directly constrained by observations. We then turn to the two alternative DM models under consideration. While the gravitational potentials of the bulge and the halo can be treated analytically due to their approximate spherical symmetry, the disk component requires a numerical approach because of its flattened, axisymmetric geometry. Before presenting the detailed properties of each mass component, we first examine the fundamental characteristics of the gravitational potential generated by a disk, starting from a few idealized models that capture its essential features.

Note that the masses of the thin and thick stellar disks are not measured directly but are
inferred within a global Galactic mass model by combining star counts,
photometric information, and dynamical constraints \citep{Juric_etal_2008}.  Observational surveys
(e.g.\ \textit{Gaia} combined with spectroscopic data) are used to determine the
spatial density profiles of the disk components, typically parametrized by
exponential radial scale lengths and exponential or $\mathrm{sech}^2$ vertical
profiles.  Local normalizations are fixed by star counts in the solar
neighborhood, while stellar population synthesis models are employed to convert
number densities and luminosities into mass densities via assumed initial mass
functions and mass--to--light ratios.  The resulting density models are then
embedded in a self-consistent gravitational potential constrained by the
Galactic rotation curve and by vertical dynamical equilibrium through the Jeans
and Poisson equations.  The disk masses follow from integrating the inferred
mass density profiles over the Galactic volume and are therefore
model-dependent quantities, constrained primarily by the total mass
distribution rather than by the kinematics of any single tracer population.

%%%%%%%%%%%%%%%%%%%%%%%%%%%%%%%%%%%%%%%%%%%%%%%%%%%%%%%%%%%%
%%%%%%%%%%%%%%%%%%%%%%%%%%%%%%%%%%%%%%%%%%%%%%%%%%%%%%%%%%%%

\subsection{Exponential disk}

 A standard reference for the disk rotation curve is the razor thin-exponential disk, 
for which the surface mass density in the mid-plane is  
$
\Sigma(R) = \Sigma_0 \, e^{-R/R_{\rm d}}, \qquad (z=0).
$
In this case, the circular velocity in the mid-plane is given by \citep{Freeman_1970} 
\[
v_{c,{\rm disk}}^{\,2}(R, 0)
= 4 \pi G \, \Sigma_0 R_{\rm d} \, y^2 
\left[ I_0(y)K_0(y) - I_1(y)K_1(y) \right],
\]
where 
$
y \equiv \frac{R}{2R_{\rm d}},
$ 
and \(I_n\) and \(K_n\) denote the modified Bessel functions of the first and second kind, respectively.
The maximum of mid plane circular velocity  is 
\[
\frac{v_{c,{\rm disk}}^{\,2}(R_{\rm max},0)}{v_{\rm disk}} \approx 0.6 \;, 
\] 
where 
$
v_{\rm disk}^2 = \frac{G M_{\rm disk}}{R_{\rm d}} = \frac{4 \pi G \Sigma_0  R_{\rm d}^2  }{R_{\rm d}}
$ 
and is reached for
$
R_{\rm max} \approx 2 R_{\rm d} \;. 
$ 
\medskip

{   
For a disk with a double-exponential density profile
\begin{equation}
\rho(R,z) = \rho_0\, e^{-R/R_{\text{d}}}\, e^{-|z|/z_{\text{d}}},
\end{equation}
the vertical acceleration can be approximated as
\begin{equation}
\label{eq:az_expdisk}
a_z(R,z) \;\simeq\;
-4\pi G\, \rho_0\, e^{-R/R_{\text{d}}}\, z_{\text{d}}
\left[1 - e^{-|z|/z_{\text{d}}}\right].
\end{equation}
This expression has the correct asymptotic behaviour: for
$|z|\ll z_{\text{d}}$, one finds
$a_z \propto z$, while for $|z|\gg z_{\text{d}}$ the acceleration approaches the constant value expected for an infinite exponential disk with surface density
$\Sigma(R)=2\rho_0 z_{\text{d}} e^{-R/R_{\text{d}}}$.

For a fixed disk mass $M_{\text{d}}$, the dependence on the vertical scale height is significant mainly around $z\sim z_{\text{d}}$. In this regime,
\begin{equation}
\label{eq:az_approx} 
a_z(R,z) \;\simeq\;
G\,\frac{M_{\text{d}}}{R_{\text{d}}^2}\,e^{-R/R_{\text{d}}}
\left[1 - e^{-|z|/z_{\text{d}}}\right],
\end{equation}
showing explicitly that $z_{\text{d}}$ affects only the vertical transition term, while the overall amplitude scales as
$M_{\text{d}}/R_{\text{d}}^2$.
Physically, $a_z$ grows approximately linearly near the mid-plane and saturates for $|z|\gtrsim z_{\text{d}}$.
}

\subsection{Bulge} 

The bulge density profile is usually modeled with a Hernquist distribution \citep{BinneyTremaine2008},  
\be
\label{bulge_rho} 
\rho(r) \;=\; \frac{\rho_{\rm b,0} }{\left(\dfrac{r}{r_{\rm b} }\right)\left(1+\dfrac{r}{r_{\rm b} }\right)^3},
\qquad r = \sqrt{R^2+z^2},
\ee
with total mass 
$
M_{\rm b} \;=\; 2\pi \rho_{\rm b, 0} r_{\rm b}^3.
$
The enclosed mass profile is 
\[
M_{\rm b} (r) \;=\; 4\pi \!\int_0^r \rho(r')\,r'^2 \,dr'
= M_{\rm b} \,\frac{r^2}{(r+r_{\rm b})^2}.
\]

For the MW bulge, we adopt a scale radius $r_{\rm b} = 0.25$ kpc and a mass 
$M_{\rm b} = 2 \times 10^{10} M_\odot$ \citep{Juric_etal_2008}.

The corresponding gravitational potential is 
\be
\label{bulge_phi} 
\Phi_{\rm b}(r) \;=\; -\,\frac{G M_{\rm b}}{r+r_{\rm b}},
\qquad \Phi_{\rm b}(\infty)=0.
\ee
By spherical symmetry, the gravitational acceleration is purely radial,  
with mid-plane circular velocity ($z=0$) 
\[
v_{c,{\rm b}}^2(R,0) \;=\; \frac{G M_{\rm b}\,R}{(R+r_{\rm b})^2}.
\]

More generally, the circular velocity at height $z$, defined as the azimuthal speed balancing the radial force at fixed $z$, is  
\be
\label{bulge_vc} 
v_{c,{\rm b}}^2(R,z) 
= R \,\frac{\partial \Phi_{\rm b}}{\partial R}(R,z)
= \frac{G M_{\rm b}\,R^2}{r\,(r+r_{\rm b})^2} \;. 
\ee
We obtain for the vertical acceleration
\be
\label{bulge_az}
  a_{z,b}(R,z) = -\frac{d\Phi}{dr} \cdot \frac{\partial r}{\partial z} 
           = -\left( \frac{G M_b}{(r + r_b)^2} \right) \cdot \frac{z}{r} \;.
\ee
This acceleration is directed toward the mid-plane \(z = 0 \), as expected from a spherically symmetric mass distribution.

%%%%%%%%%%%%%%%%%%%%%%%%%%%%%%%%%%%%%%%%%%%%%%%%%%%%%%%%%%%%
%%%%%%%%%%%%%%%%%%%%%%%%%%%%%%%%%%%%%%%%%%%%%%%%%%%%%%%%%%%%
%%%%%%%%%%%%%%%%%%%%%%%%%%%%%%%%%%%%%%%%%%%%%%%%%%%%%%%%%%%%

\subsection{Thin disk} 

A simple and widely used three-dimensional model for the thin disk mass density is the double-exponential function
with $\rho_{{ \rm tnd},0}$ as the normalization constant and $R_{{ \rm tnd}}$ and $z_{{ \rm tnd}}$ as the characteristic scales for the exponential decay respectively in the radial and vertical directions.  In the limit $R_{{ \rm tnd}} \gg z_{{ \rm tnd}}$, the razor thin approximation provides a good description of the mid-plane rotation curve.

Typical parameters of the thin disk are $R_{ \rm tnd} = 4.5$ kpc and a mass 
$M_{ \rm tnd} = 3 \times 10^{10} \, M_\odot$ 
\citep{Eilers_etal_2019,SylosLabini_etal_2023}.  
The vertical scale height is $z_{{ \rm tnd}} = 0.14$ kpc, so that the central density is
\be
 \rho_{{ \rm tnd},0} = \frac{M_{\text{{ \rm tnd}}} }{4 \pi R_{{ \rm tnd}}^2 z_{{ \rm tnd}}} 
 \;\approx\; 8.4 \times 10^{8} \; M_\odot \, \text{kpc}^{-3}.
\ee

The rotation velocity of the think disk is found by making the numerical derivatives of the potential $\Phi_{\rm tnd}$ which is obtained by using the Poisson solver
(see Appendix A).
%%%%%%%%%%%%%%%%%%%%%%%%%%%%%%%%%%%%%%%%%%%%%%%%%%%%%%%%%%%%
%%%%%%%%%%%%%%%%%%%%%%%%%%%%%%%%%%%%%%%%%%%%%%%%%%%%%%%%%%%%
%%%%%%%%%%%%%%%%%%%%%%%%%%%%%%%%%%%%%%%%%%%%%%%%%%%%%%%%%%%%

\subsection{Thick disk} 

The density of the thick disk can be modeled with a double-exponential profile 
with $\rho_{{ \rm tkd},0}$ as the normalization constant and $R_{{ \rm tkd}}$ and $z_{{ \rm tkd}}$ as the characteristic scales for the exponential decay respectively in the radial and vertical directions.

The total mass of the thick disk is \citep{Eilers_etal_2019,SylosLabini_etal_2023}  
%\[
$M_{\text{{ \rm tkd}}} = 2.7 \times 10^{10} \, M_\odot$,
%\]
the radial scale length is 
%\[
$R_{{ \rm tkd}} = 2.3 \, \text{kpc}$,
%\
and the vertical scale height %is 
%\[
$z_{{ \rm tkd}} = 1.21 \, \text{kpc}$.
%\]
This implies a central density 
\be
\rho_{{ \rm tkd},0} = \frac{M_{\text{{ \rm tkd}}}}{4 \pi R_{{ \rm tkd}}^2 z_{{ \rm tkd}}} 
= 3.4 \times 10^{8} \, M_\odot \, \text{kpc}^{-3}.
\ee

%Figure~\ref{vc_thick_disk} (upper panel) shows 

{    
The generalized rotation curve at different heights $z$ and 
 the vertical acceleration are similar to that of the thin disk but for a rescaling due to a different total mass and 
different characteristic length scales.
Note that, unlike the thin disk case, the exponential thin-disk approximation with parameters $R_{\text{tkd}}$ and $M_{\text{tkd}}$ does not provide a good description of the mid-plane rotation curve.  
This discrepancy arises because, for the thick disk, the vertical scale height is comparable to (about half of) the radial scale length, so vertical structure plays a significant role in shaping the rotation curve.  

As for the case of the thin disk, the vertical variation of both the radial and vertical gravitational accelerations is strongest in the range \( R \sim R_{\text{d}}  \),  i.e., at about the 
characteristic scale length of the stellar disk. This enhances the sensitivity of dynamical diagnostics in this region.
}

%%%%%%%%%%%%%%%%%%%%%%%%%%%%%%%%%%%%%%%%%%%%%%%%%%%%%%%%%%%%
%%%%%%%%%%%%%%%%%%%%%%%%%%%%%%%%%%%%%%%%%%%%%%%%%%%%%%%%%%%%
%%%%%%%%%%%%%%%%%%%%%%%%%%%%%%%%%%%%%%%%%%%%%%%%%%%%%%%%%%%%

\subsection{The neutral hydrogen disk} 

The density profile of the gaseous disk is assumed to follow that of the neutral hydrogen (\HI), which is the dominant gas component. 
According to \citet{Kalberla_Dedes_2008}, the distribution can be approximated by a double exponential
with parameters are  
%\[
$M_{{\rm H \textsc{i}}} \;=\; 0.5 \times 10^{10} \, M_\odot, 
\qquad R_{{\rm H \textsc{i}}}=9.8 \,\text{kpc}, 
\qquad z_{{\rm H \textsc{i}}}=0.15 \,\text{kpc},
$
%\]
which yield a normalization  
\be
 \rho_{{\rm H \textsc{i}},0} \;=\; \frac{M_{{\rm H \textsc{i}}}}{4 \pi R_{{\rm H \textsc{i}}}^2 z_{{\rm H \textsc{i}}}} 
 \;=\; 2.8 \times 10^{7} \, M_\odot \,\text{kpc}^{-3}.
\ee
 Note that in this case, the agreement with the exponential disk approximation in the mid-plane is essentially exact, since
  $R_{{\rm H \textsc{i}}} \approx 60  z_{{\rm H \textsc{i}}}$.

%%%%%%%%%%%%%%%%%%%%%%%%%%%%%%%%%%%%%%%%%%%%%%%%%%%%%%%%%%%%
%%%%%%%%%%%%%%%%%%%%%%%%%%%%%%%%%%%%%%%%%%%%%%%%%%%%%%%%%%%%
%%%%%%%%%%%%%%%%%%%%%%%%%%%%%%%%%%%%%%%%%%%%%%%%%%%%%%%%%%%%
%%%%%%%%%%%%%%%%%%%%%%%%%%%%%%%%%%%%%%%%%%%%%%%%%%%%%%%%%%%%
%%%%%%%%%%%%%%%%%%%%%%%%%%%%%%%%%%%%%%%%%%%%%%%%%%%%%%%%%%%%
%%%%%%%%%%%%%%%%%%%%%%%%%%%%%%%%%%%%%%%%%%%%%%%%%%%%%%%%%%%%

\subsection{The dark matter halo} 

The density profile of a Navarro-Frank-White (NFW) halo is defined by two free parameters, the characteristic density $\rho_{\rm h}$ and scale radius $r_{\rm h}$:
\[
\rho_{{\rm h}}(r) \;=\; \frac{\rho_{\rm h}}{x(1+x)^2}, 
\qquad 
x \equiv \frac{r}{r_{\rm h}}, 
\qquad 
r=\sqrt{R^2+z^2}.
\]

The enclosed mass profile is 
\[
M_{{\rm h}}(r) \;=\; 4\pi \rho_{\rm h} r_{\rm h}^3 
\left[ \ln(1+x) - \frac{x}{1+x} \right],
\]
and the gravitational potential is 
\[
\Phi_{{\rm h}}(r) \;=\; -\,4\pi G\,\rho_{\rm h}\,r_{\rm h}^2 \,\frac{\ln(1+x)}{x}.
\]

In cylindrical coordinates, the force components are 
\[
\frac{\partial \Phi_{{\rm h}}}{\partial R}
= \frac{d\Phi_{{\rm h}}}{dr}\,\frac{R}{r}
= \frac{G\,M(r)}{r^2}\,\frac{R}{r},
\]
and 
\[
\frac{\partial \Phi_{{\rm h}}}{\partial z}
= \frac{d\Phi_{{\rm h}}}{dr}\,\frac{z}{r}
= \frac{G\,M(r)}{r^2}\,\frac{z}{r}.
\]

The generalized circular velocity, defined as the azimuthal speed that balances the radial force at fixed $z$, is 
\be
\label{halo_vc} 
v_{{\rm c,h}}^2(R,z) \;\equiv\; R\,\frac{\partial\Phi_{{\rm h}}}{\partial R}(R,z)
= \frac{G\,M(r)\,R^2}{r^3},
\ee

In the mid-plane ($z=0$) this reduces to 
\bea
&&
v_{{\rm c,h}}^2(R,0) \;=\; \frac{G\,M(R)}{R}
\\ \nonumber &&
= 4\pi G\,\rho_{\rm h} r_{\rm h}^3 
\,\frac{\ln(1+R/r_{\rm h}) - \dfrac{R/r_{\rm h}}{1+R/r_{\rm h}}}{R}.
\eea

The central potential value is finite:
\[
\Phi_{{\rm h}}(0) \;=\; -\,4\pi G\,\rho_{\rm h}\,r_{\rm h}^2,
\qquad
\text{since } \lim_{x\to 0}\frac{\ln(1+x)}{x}=1.
\]

The vertical acceleration \( a_{z,h}(R, z) \) is defined as the vertical gradient of the gravitational potential:
\bea
&&
a_{z,h}(R, z)  = \frac{\partial \Phi}{\partial z} = \frac{d\Phi}{dr} \cdot \frac{\partial r}{\partial z} = \frac{d\Phi}{dr} \cdot \frac{z}{r} \;\; 
\\ \nonumber 
&&
\label{halo_az} 
= - \frac{4\pi G \rho_h r_h^2  z}{r^3} \left[ \ln\left(1 + \frac{r}{r_h} \right) - \frac{r/r_h}{1 + r/r_h} \right],
\eea
This expression can be used to compute the vertical component of the gravitational acceleration due to a NFW halo at any cylindrical position \( (R, z) \).

%%%%%%%%%%%%%%%%%%%%%%%%%%%%%%%%%%%%%%%%%%%%%%%%%%%%%%%%%%%%
%%%%%%%%%%%%%%%%%%%%%%%%%%%%%%%%%%%%%%%%%%%%%%%%%%%%%%%%%%%%
%%%%%%%%%%%%%%%%%%%%%%%%%%%%%%%%%%%%%%%%%%%%%%%%%%%%%%%%%%%%

\subsection{The dark matter disk}

The density of the DM disk is also  modeled with a double-exponential profile
where the  radial scale length $R_{{ \rm m}}$  and the  mass 
%\[
$M_{{ \rm m}} = 4 \pi \rho_{{ \rm m},0} R_{{ \rm m}}^2 z_{{ \rm m}}$ 
%\]
are the free parameters of the model. In principle, also the vertical scale height $z_{{ \rm m}}$ can be a free parameter. 
In particular in what follows we will fix $z_{{ \rm m}}$ and let only $R_{{ \rm m}}, M_{{ \rm m}}$ be free 
parameters. 
%By considering 
{    
The values obtained by \cite{SylosLabini_etal_2023} for the mid-plane are $R_{\rm m} = 8\,\mathrm{kpc}$ and $M_{\rm m} = 11\,M_\odot$, with $z_{\rm m} = 0.15\,\mathrm{kpc}$. These values serve as a useful reference for the expected order of magnitude of the fitted parameters in the present analysis.
}

\subsection{Multicomponent Jeans equation}

In our model 
the density of the disc can be approximated as the sum of $N_c$ different contributions approximately double-exponentially decaying with different characteristic scales, we have 
\be
\label{multi1}
\rho_\text{\text{}}(R,z)
= \sum_i^{N_c} \rho^i_0(R,z) \exp\left(-\frac{R}{R_{{\rm d,}i}}  \right) \exp\left(-\frac{z}{z_{{\rm d,}i}} \right)
\ee 
where $R_{{\rm d,}i}$ and $z_{{\rm d,}i}$ are respectively the characteristic radial and vertical length scales of the $i^{th}$ component.  
Given the limited range of radial and vertical distances covered by our sample we neglect any radial and/or vertical dependence of
$R_{{\rm d,}i}$ and/or $z_{{\rm d,}i}$.

The fundamental assumption underlying the DM disk  model is that all mass components of the disk --- namely the stellar, gaseous, and DM components --- move within the mean gravitational potential generated by the total mass distribution, that is, the sum of the stellar, gaseous, and DM contributions, with the latter assumed to have a flattened disk-like geometry.
In contrast, in the standard disk-halo model, the system is also governed by the total potential of stars, gas, and DM, but in this case the DM is distributed in a quasi-spherical halo.
While the baryonic components in the disk move coherently within this potential --- consistently with the mean-field approximation --- the DM particles move within the halo potential and therefore do not exhibit a net circular motion. Instead, they are characterized by an almost isotropic velocity dispersion.

In summary, the two models differ in their geometrical and kinematical assumptions regarding the DM distribution: in the DMD model, DM is flattened and co-rotates with the baryonic disk, whereas in the halo model it forms a nearly spherical, dynamically hot component supported by random motions.

By defining the quantities 
\bea
&&
\label{h1} 
\overline{ R_{d}}   = - \left(\frac{1} {\rho}  \frac{\partial \rho} { \partial R } \right)^{-1} 
\\ \nonumber && 
\overline{ z_{d}}   = - \left(\frac{1} {\rho }  \frac{\partial \rho}  { \partial z } \right)^{-1} \;.
\eea
 we can rewrite the Jeans equations  Eq.\ref{eq:Jeans_eff} as  
\bea
\label{eq:Jeans_eff2}
&&
v_c^2(R,z) 
\approx \overline{v_\phi}^2 - {\sigma^2_{\phi}} + 
{\sigma^2_{R}} \left( 1 + \frac{R}{ \overline{ R_{d}} } +   \frac{\partial \ln \sigma^2_{R}} { \partial \ln R }  \right) 
\\ \nonumber && 
a_z(R,z)  
\approx   \frac{\text{sign}(z)} { \overline{ z_{d}} }\sigma^2_{z} - \frac{\partial  {\sigma_{z}^2} }  { \partial z}   \;, 
\eea

From Eq.\ref{eq:Jeans_eff2} it follows that  both the determinations of $v_c$ and $a_z$ from the kinematic data require the knowledge of the density distribution of the disk, as the quantities $\overline{R_{{\rm d}}}$ and $\overline{ z_{{\rm d}}}$ are involved on the right-hand side of Eq.\ref{eq:Jeans_eff2}, respectively. However, the circular velocity is primarily influenced by the tangential velocity $\overline{v^2_\theta}$, and the terms involving $\overline{R_{{\rm d}}}$ represent only second-order perturbations. On the other hand, the vertical acceleration strongly depends on $\overline{z_{{\rm d}}}$, as it appears in the denominator of the second equation in  Eq.\ref{eq:Jeans_eff2}.

If we consider the stellar and \HI\ components and neglecting the dependences on $R,z$ of all length scales we find 
\bea
\label{h2} 
&&
\overline{ R_{{\rm d}}}  = 3.5 \;\; \text{kpc}  
\\ \nonumber && 
\overline{ z_{{\rm d}}}  =  0.19 \;\; \text{kpc}  \;. 
\eea
Concerning the uncertainties in the determination of the vertical scale heights of the stellar components, 
we adopt the representative value $\overline{z_{{\rm d}}} = 0.20 \pm 0.02$~kpc, corresponding to an uncertainty of about 10\%.

If we also add the DM disk (the best fit parameters obtained below for $M_{{\rm m}}$ and $R_{{\rm m}}$ but assuming $z_{{\rm m}}=0.15(1.0)$ kpc) 
\bea
\label{h3} 
&&
\overline{ R_{d}}  = 4.7 \;\; \text{kpc}  
\\ \nonumber && 
\overline{ z_{d}}  =  0.17 (0.27) \;\; \text{kpc}  \;.
\eea
Given that the vertical scale height of the DM disk is a free parameter of the model, 
we adopt the value $\overline{z_{{\rm d}}} = 0.22 \pm 0.04$~kpc, corresponding to a larger uncertainty 
of about 20\% compared to the 10\% adopted for the stellar disk.

\subsection{Discussion} 

\begin{figure} 
\includegraphics[width=0.45\textwidth]{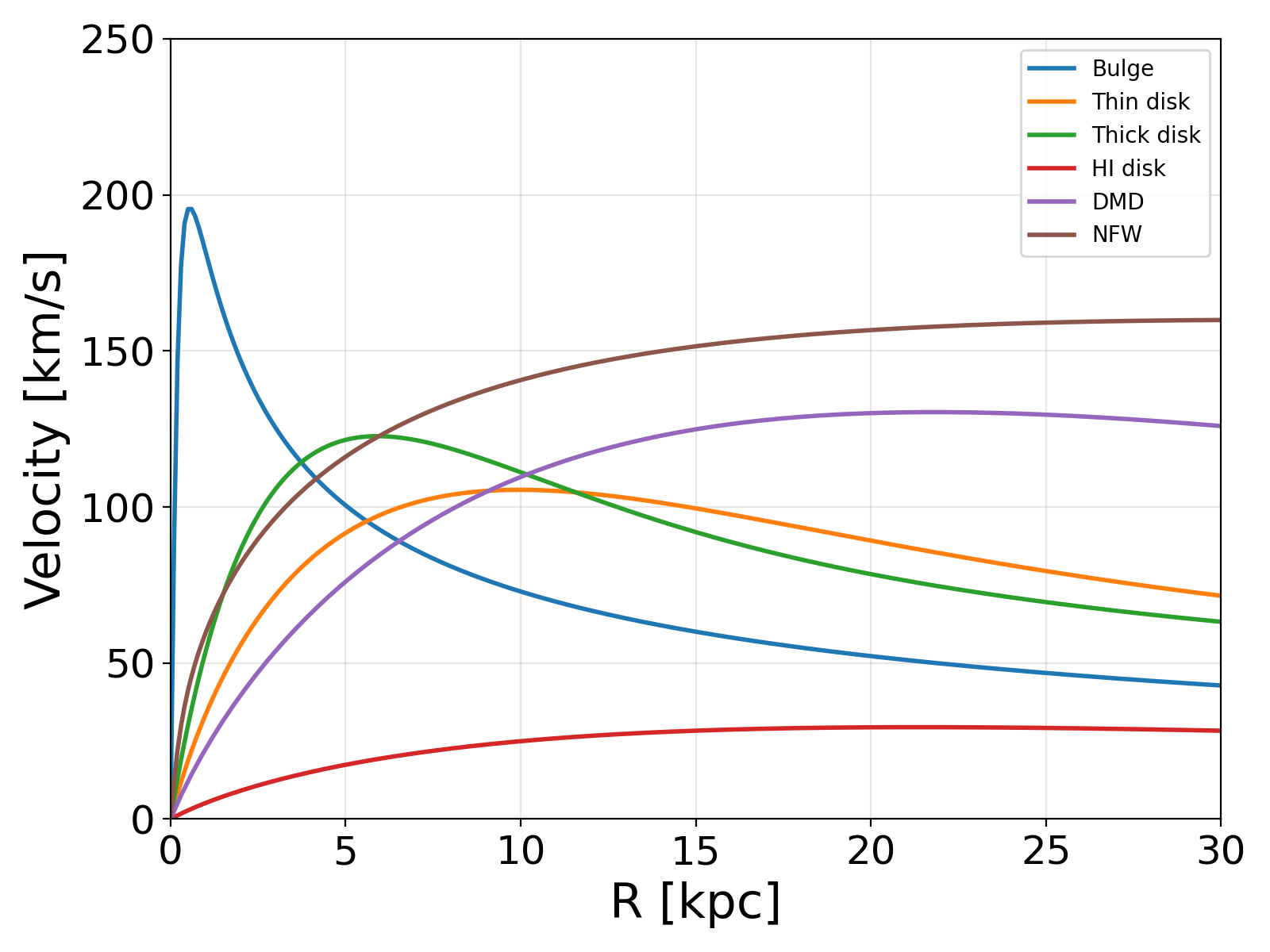}
\includegraphics[width=0.45\textwidth]{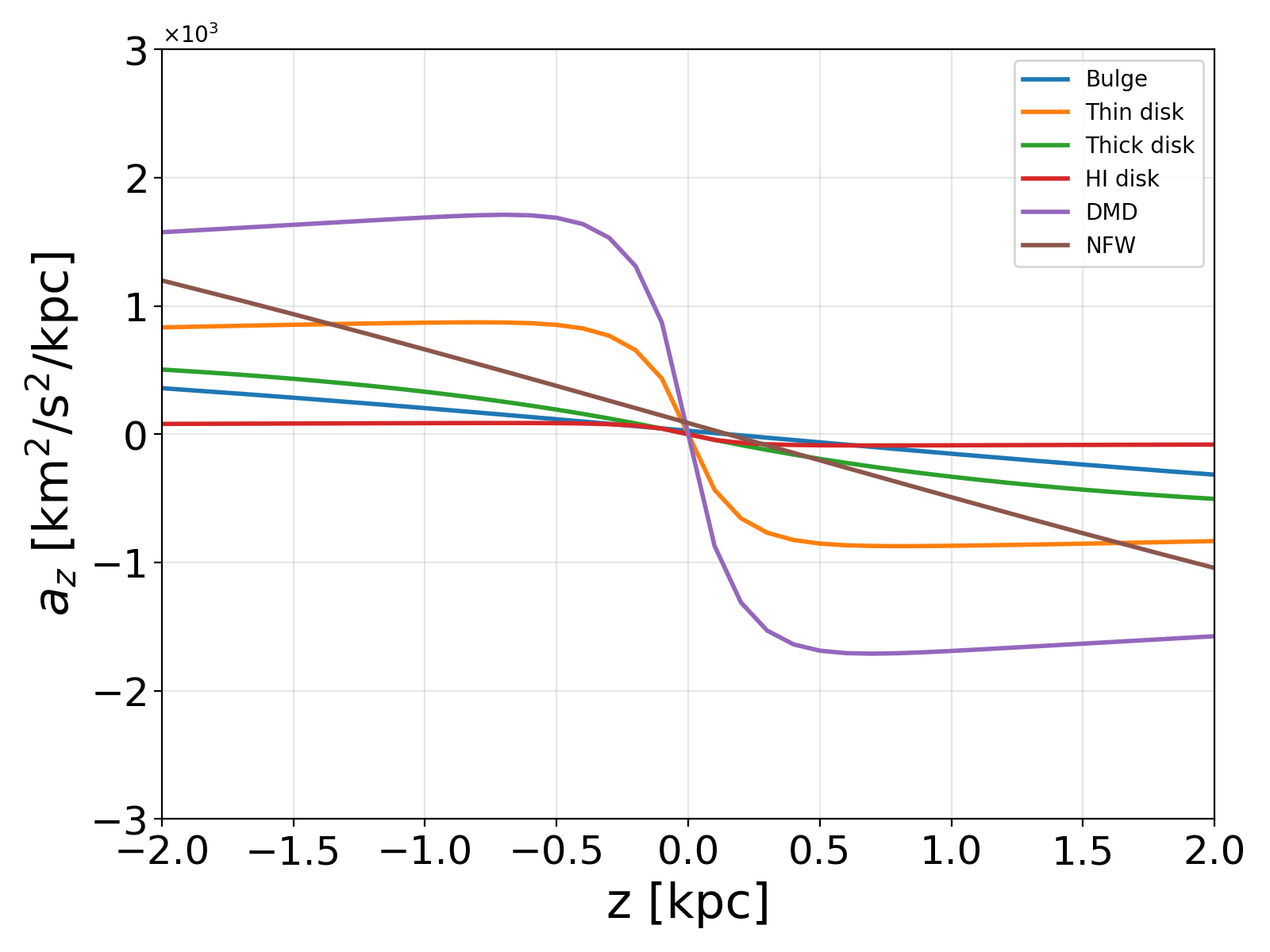}
\caption{
(i) Contributions to the mid-plane rotation curve $v_c(R,z=0 \; \text{kpc})$ of the stellar components, the NFW halo
 and DM disk. 
(ii) Contributions to the vertical acceleration for $a_z(R=9\; \text{kpc},z)$ of the stellar components, the NFW halo
 and DM disk with $z_{\text{d} }=0.15$ kpc. 
 } 
\label{fig:Vc_z0} 
\end{figure}

Fig.~\ref{fig:Vc_z0} shows the contribution of the different stellar components together with the two DM models, i.e. the NFW halo and the DMD model, for $z=0$ and the same parameters adopted above. The difference between the two models becomes particularly evident at large radii ($R \gtrsim 15$ kpc), where the DMD curve declines more rapidly than the halo model. 

The strongest differences between rotation curves at different vertical heights arise in the inner disk. Around the radius of the mid-plane maximum ($\sim 2 R_{\text{d}}$, with $R_{\text{d}}$ the exponential scale length), the rotation speed shows a marked dependence on $z$. By contrast, in the case of the DM halo the variation with vertical height is almost negligible, reflecting its spherical symmetry: with a typical scale radius of $R_{{\rm s}} \approx 15$ kpc, changes in $z$ barely modify its gravitational contribution.

This distinction has important implications for mass modeling. In standard NFW-based models, where the total potential is the sum of a spherical halo and the stellar components, the stellar contribution is observationally fixed and the two halo parameters have little impact on $v_c(R,z)$. In the DM disk scenario, instead, the total potential combines the stellar components with a DM disk. Here the free parameters of the dark disk significantly alter the generalized rotation curves, leading to a much stronger dependence on vertical height.

\subsection{Inverse Poisson Equation}

The method discussed above and adopted in this work is based on comparing the generalized rotation curves 
$v_c(R,z)$ and the vertical acceleration $a_z(R,z)$, as estimated from the data via the Jeans equations, 
with those predicted by a given mass model through the derivatives of the corresponding gravitational potential. 
This approach allows us to isolate the role of the model parameters, and,  in particular, for the DM disk model, 
to investigate the impact of its two characteristic length scales, $R_{{\rm m}}$ and $z_{{\rm m}}$, whose influence on the radial and 
vertical terms is distinct. 

On the other hand, several authors 
\citep[e.g.][]{MoniBidin_etal_2012a,MoniBidin_etal_2012b,Lopez-Corredoira_2025} 
have adopted an alternative approach to compare observational data with mass models. 
Rather than directly analysing the acceleration fields, these studies estimate the local mass density from 
kinematic quantities by inverting the Poisson equation, which relates the gravitational potential $\Phi$ to 
the mass density $\rho$ through
\begin{equation}
\label{eq:poisson2}
\nabla^2 \Phi = 4\pi G \rho \,.
\end{equation}
For a given density distribution $\rho(R,z)$, this equation can be solved to determine the 
corresponding gravitational potential $\Phi(R,z)$. 
Conversely, the \emph{inverse problem} consists in recovering the underlying density 
$\rho(R,z)$ from an estimated potential $\Phi(R,z)$, inferred from the observed kinematic 
quantities by means of the Jeans equations, under the standard assumptions of axisymmetry 
(i.e. independence of the azimuthal coordinate $\varphi$) and that both $\Phi$ and $\rho$ 
depend only on the cylindrical coordinates $(R,z)$.
{    
This inversion is known to be a classic ill-posed inverse problem, usually overcome by authors by applying standard techniques to recover solutions that are sufficiently stable and reliable for their purposes. Note that this potential problem is avoided with the direct approach adopted in this paper.
}

By defining
\begin{equation}
\label{eq:sigma}
\Sigma(R,z) \equiv \int_{-z}^{z} \rho(R,y)\,dy,
\end{equation}
and combining Eq.~\ref{eq:poisson2} with the Jeans equations 
(Eqs.~\ref{eq:jeansR_full}–\ref{eq:jeansZ_full}), one obtains
\bea
\label{eq:sigma_est}
&&
\Sigma(R,z) = \frac{1}{4\pi G}\,\frac{1}{R}
\int_{-z}^{z} \frac{\partial v_c^2(R,y)}{\partial R}\,dy 
\\ \nonumber && + \left[a_z(R,z) - a_z(R,-Z)\right].
\eea

To estimate $\Sigma(R,z)$, one must therefore determine not only the circular velocity 
$v_c(R,z)$ and the vertical acceleration $a_z(R,z)$, but also the integral of the radial derivative 
of $v_c^2(R,z)$. 
Hence, $\Sigma(R,z)$ depends on a combination of both the radial and vertical Jeans equations. 
As discussed in Sect.~\ref{sect:data}, while the circular velocity is, to first order, determined by 
the tangential velocity, the vertical acceleration is strongly influenced by the 
characteristic vertical scale length of the DM disk component. 
For this reason, in the present work we have adopted an approach based on the 
independent estimation of the radial and vertical accelerations. 
Naturally, if a satisfactory fit is obtained for both $v_c(R,z)$ and $a_z(R,z)$, 
Eq.~\ref{eq:sigma_est} implies that --- neglecting noise and small fluctuations arising from the 
radial derivative --- a similarly good agreement should be achieved for $\Sigma(R,z)$.

%%%%%%%%%%%%%%%%%%%%%%%%%%%%%%%%%%%%%%%%
%%%%%%%%%%%%%%%%%%%%%%%%%%%%%%%%%%%%%%%%
%%%%%%%%%%%%%%%%%%%%%%%%%%%%%%%%%%%%%%%%

\section{Results with the Gaia DR3 data}
\label{sect:results} 

We adopted a constant velocity uncertainty of $\Delta v_\phi = 3~\mathrm{km\,s^{-1}}$ across the entire radial range of the sample. 
Under this assumption, the best-fitting dark matter disk (DMD) model is characterized by 
$R_{\mathrm{m}} = 7.0~\mathrm{kpc}$ and $M_{\mathrm{m}} = 1.0 \times 10^{11}~M_\odot$, i.e. about the mass of the stellar component, 
yielding a total Galactic mass of $M_{\mathrm{MW}} = 1.8 \times 10^{11}~M_\odot$.

Notably, the best-fit scale radius of the DM disk, $R_{\mathrm{m}}$, closely matches that of the \HI disk, 
i.e., $R_{\mathrm{m}} \approx R_{\text{\HI}}$. This supports the idea that, within this model, 
the distribution of neutral hydrogen gas serves as a reliable tracer of the underlying DM distribution.

For comparison, the best-fitting NFW halo model corresponds to 
$\rho_{\mathrm{s}} = 7.0 \times 10^{6}~M_\odot\,\mathrm{kpc^{-3}}$ and 
$r_{\mathrm{s}} = 18~\mathrm{kpc}$, 
which translate into a virial mass of $M_{\mathrm{vir}} = 8.0 \times 10^{11}~M_\odot$ ( i.e. about ten times the mass of the stellar component)
and a virial radius of $R_{\mathrm{vir}} = 197~\mathrm{kpc}$. 
This fit yields a total Galactic mass of $M_{\mathrm{MW}} = 8.8 \times 10^{11}~M_\odot$, i.e. 5 times larger than the DM disk mass. 

The corresponding $\chi^2$ values are reported in Table~\ref{table:chi2}, 
showing that the DMD model provides a consistently better fit to the observational data 
than the NFW halo model for all assumed uncertainties on the mean tangential velocity, $\delta v_c$.

As noted in Eq.~\ref{eq:az_approx}, for a fixed disk mass $M_{\text{d}} $, variations in the vertical scale height $z_{\text{d}} $ 
affect the vertical acceleration primarily in the vicinity of $z \sim z_{\text{d}} $, while the overall amplitude scales as $M_{\text{d}}  / R_{\text{d}} ^2$. 
This implies that the vertical scale height plays a crucial role in determining the vertical acceleration from the kinematic data through the vertical Jeans equation, 
as also shown by Eq.~\ref{eq:az_full_error_rel} and illustrated in Fig.~\ref{fig:vertical_acceleration_1}. 
Consequently, the parameter $z_{\text{d}} $ has only a marginal influence on the modeling of the DM disk itself, 
but it is critical when deriving $a_z$ from the observational data via the vertical Jeans equation. 
Naturally, the value of $z_{\text{d}} $ used in the Jeans analysis must be consistent with that adopted in the theoretical mass models. 
This requirement makes it particularly difficult to reconcile the observations with the NFW model, 
since in that case the effective vertical scale height $\overline{z_{\text{d}} } \approx 0.20 \pm 0.02 $~kpc is determined by the weighted average defined in Eq.~\ref{h2}. 
For this value of $\overline{z_{\text{d}}} $, no plausible combination of halo parameters $(\rho_{\text{s}} , r_{\text{s}} )$ yields a satisfactory fit 
to the observed vertical acceleration. 

In contrast, adopting the value $\overline{z_{\text{d}}} \approx 0.22 \pm 0.04$~kpc inferred for the DM disk model 
(see Eq.~\ref{h3}) yields a significantly better agreement with the data. 
This improvement arises because the contribution of the DM disk to the vertical acceleration 
is substantially larger than that of a spherical halo, whose effect is comparatively weak within the inner disk regions
as shown by Fig.\ref{fig:Vc_z0}.

\begin{figure} 
\includegraphics[width=0.45\textwidth]{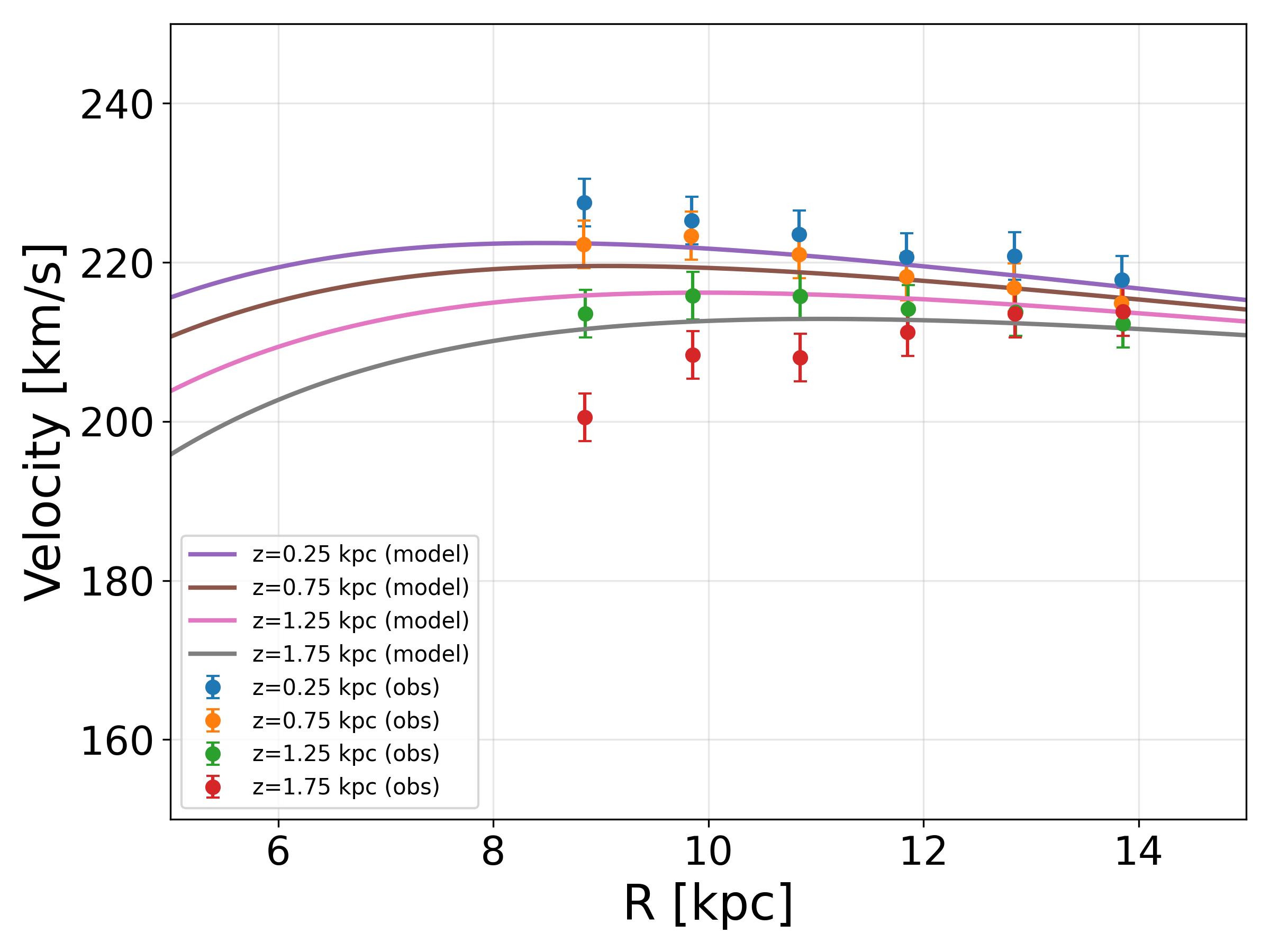}
\includegraphics[width=0.45\textwidth]{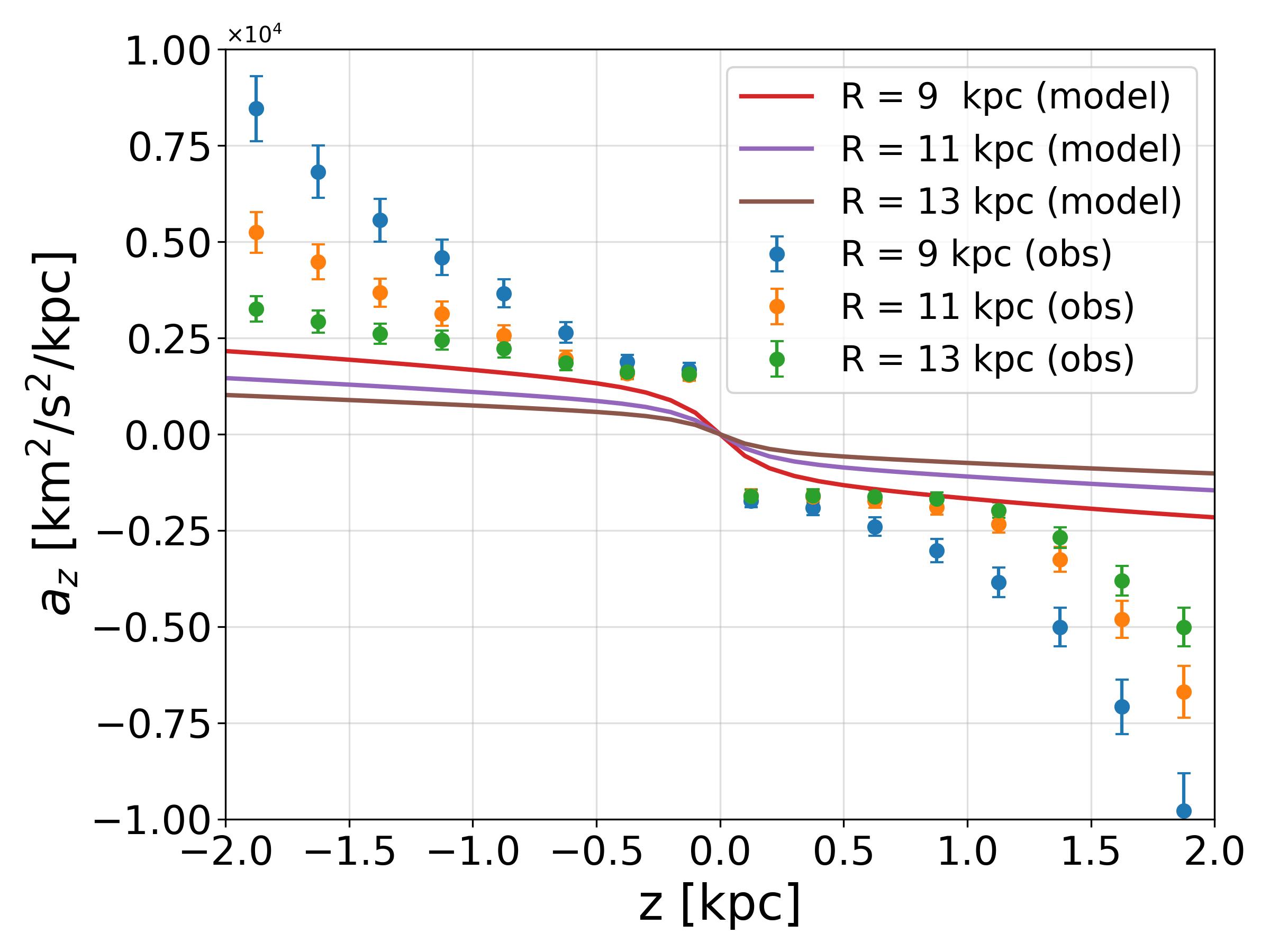}
\caption{Best fit of the NFW model to the rotation curves (top panel) and corresponding behavior of the 
vertical acceleration (bottom panel), computed assuming a mean vertical scale height 
of $\overline{z_{\text{d}}} = 0.20 \pm 0.02$~kpc.} 
\label{fig:results_nfw} 
\end{figure}

\begin{figure} 
\includegraphics[width=0.45\textwidth]{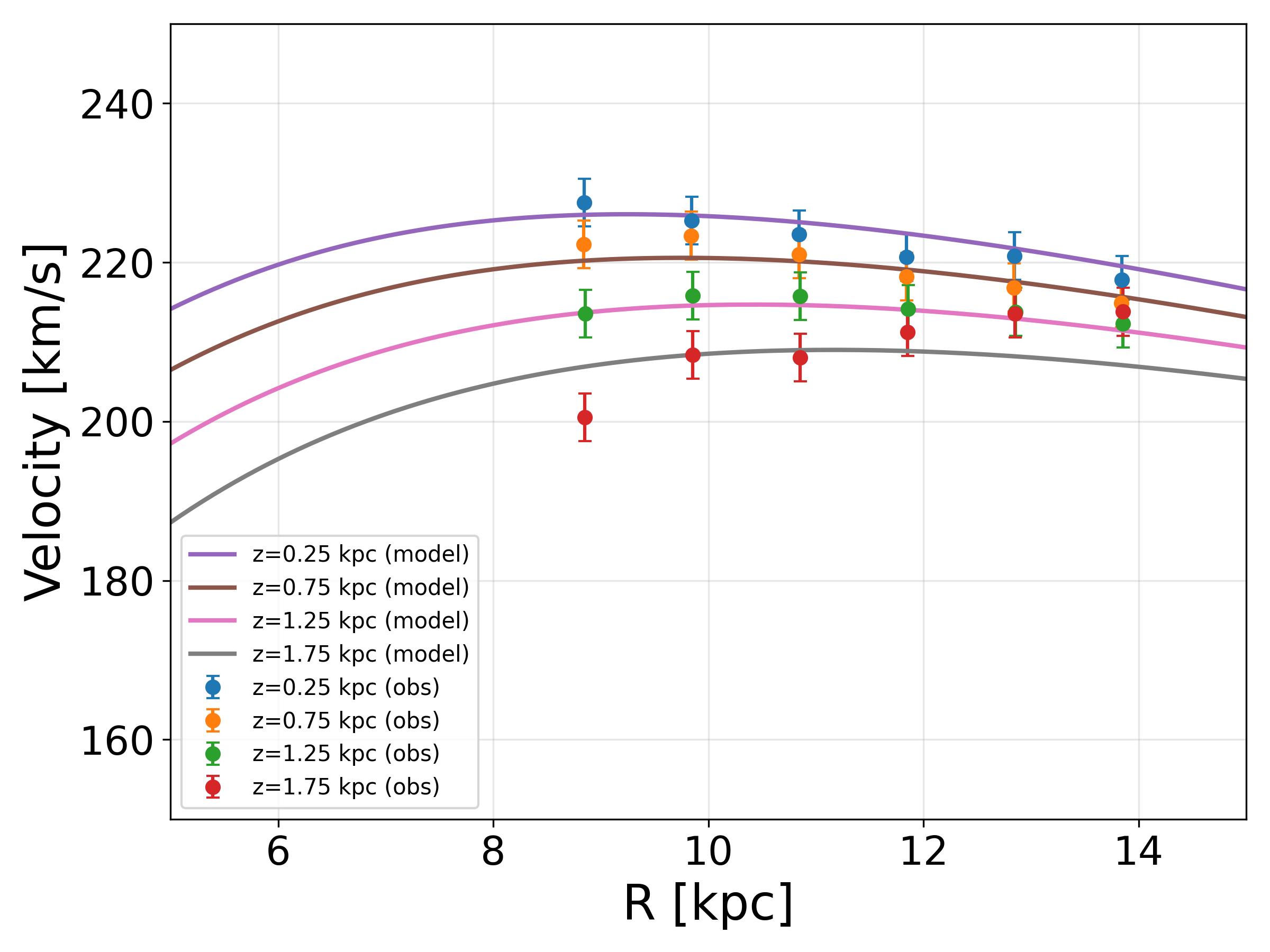}
\includegraphics[width=0.45\textwidth]{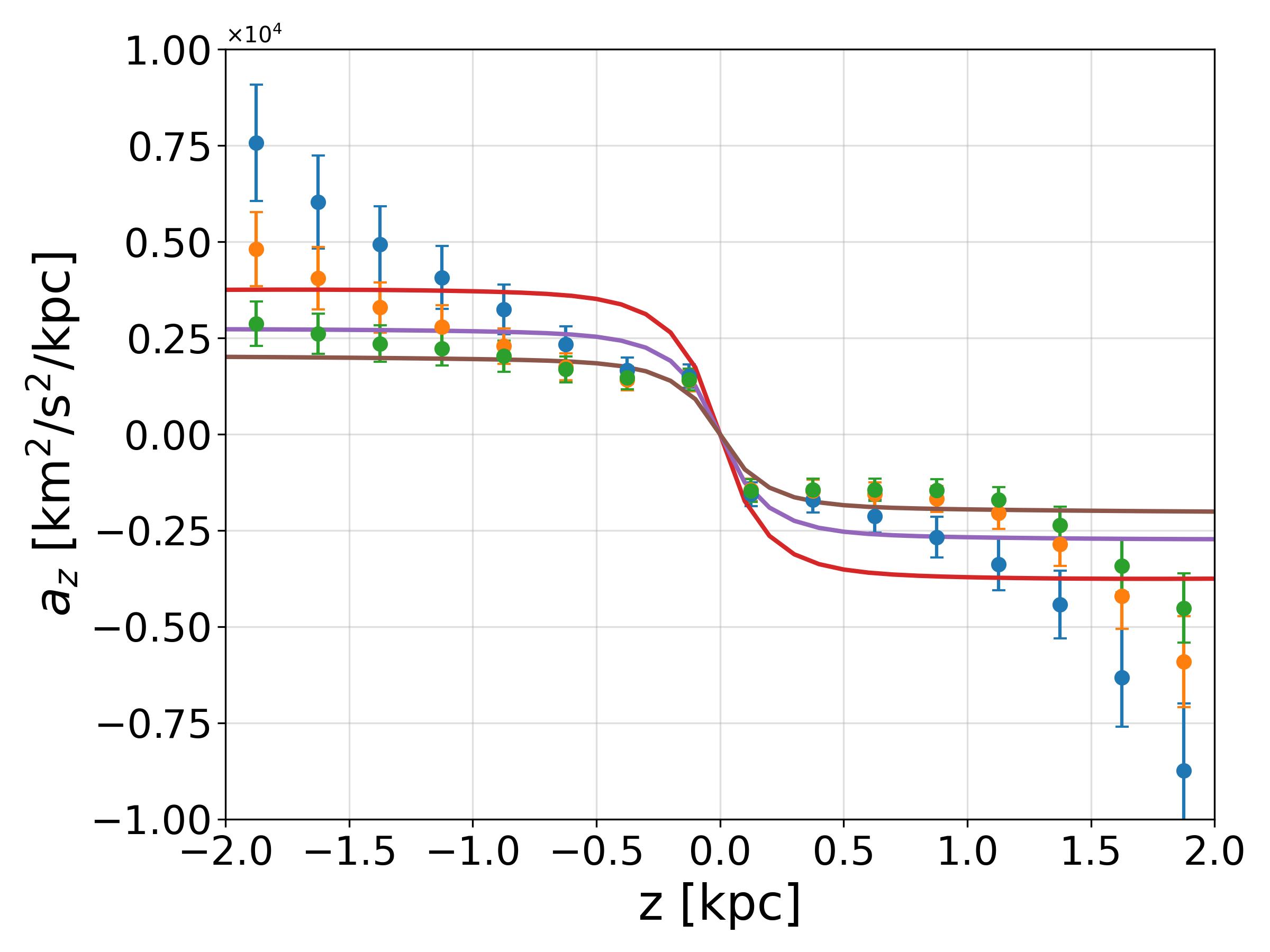}
\caption{Best fit of the DM disk model to the rotation curves (top panel) and corresponding behavior of the 
vertical acceleration (bottom panel), computed assuming a mean vertical scale height 
of $\overline{z_{\text{d}}} = 0.22 \pm 0.04$~kpc.
} 
\label{fig:results_dmd} 
\end{figure}

\begin{table}[ht] 
\caption{Reduced $\chi^2$ values for the NFW and DM disk models} 
\centering      
\begin{tabular}{|c |c |c |}  
\hline
$\delta v_c$  (km s$^{-1}$) & $\chi^2$(NFW) & $\chi^2$(DMD) \\
\hline
1		     	     &	     26.0		 & 15.0 \\
2		     	     &	     6.6		 & 4.0 \\
3		     	     &	     1.3		 & 0.8 \\
4		     	     &	     0.8		 & 0.4 \\
5		     	     &	     0.5		 & 0.3 \\
\hline                       
 \end{tabular}
  \label{table:chi2} 
   \end{table}                   
By using the best-fit parameters obtained from the minimization of the $\chi^2$ on the rotation curve data, we can also evaluate the corresponding $\chi^2$ for the vertical acceleration (bottom panels of Figs.~\ref{fig:results_nfw} and \ref{fig:results_dmd}).
Even in this case, the fit with the DM disk model provides better agreement with the data than the NFW profile, yielding $\chi^2 = 1.8$ compared to $\chi^2 = 3.3$.
This fit could be further improved by introducing an additional free parameter related to the vertical scale height or profile. However, we chose not to do so, as we consider the main constraint to come from the rotation curve, for the reasons explained above.

{
\section{Observational evidence for cold and CO--dark molecular gas}
\label{sec:cold_H2}

The possibility that a substantial fraction of the molecular gas in galaxies may reside in a cold, weakly emitting phase was first articulated in a systematic way by \citet{PfennigerCombes1994}. 
In this picture, molecular hydrogen may be organized in a highly clumpy and possibly fractal medium, characterized by low temperatures and densities such that standard tracers, in particular the CO rotational lines, become inefficient or entirely ineffective. 
Although the hypothesis was originally motivated by dynamical considerations, a growing body of observational work over the past two decades has provided indirect and, more recently, increasingly direct evidence for the existence of a significant molecular component not traced by CO emission.

A robust line of indirect evidence comes from comparisons between total gas column densities inferred from dust emission and those obtained by combining \HI{} 21--cm and CO line observations.
Far--infrared and submillimeter dust maps, from \textit{Planck} and \textit{Herschel}, consistently indicate excess column densities that cannot be accounted for by the observed atomic and CO--bright molecular gas alone \citep{Planck2011,Planck2015}.
Similarly, analyses of diffuse $\gamma$--ray emission produced by cosmic--ray interactions with interstellar gas reveal a comparable excess, often referred to as ``dark gas'', naturally interpreted as a mixture dominated by CO--dark H$_2$ with a smaller contribution from optically thick \HI{} \citep{Grenier2005,Ackermann2012}.

Further insight has been gained from detailed studies of the \HI{}--to--H$_2$ transition in diffuse and translucent clouds.
By combining dust--based column densities with \HI{} emission and absorption data, these works show that a substantial fraction of molecular gas resides in regimes where CO is photodissociated or sub--thermally excited and therefore remains undetected in standard surveys \citep{Reach2017}.
This ``CO--dark'' phase appears to be particularly relevant in low--metallicity environments and in the outer regions of galactic disks, but it is also present in the solar neighborhood.

More recently, progress toward more direct detection of such gas has been achieved through observations of ultraviolet fluorescence from molecular hydrogen.
\citet{Eos2025} reported the identification of a nearby molecular cloud detected primarily via far--UV H$_2$ emission, while remaining faint or undetectable in CO.
This result provides direct observational confirmation that cold molecular clouds can exist with little or no CO emission, lending strong support to the long--standing inference drawn from dust and $\gamma$--ray analyses.

Cold molecular structures have also been identified through large--scale surveys of cold dust emission.
\textit{Planck} has revealed numerous very cold clumps throughout the Milky Way, many of which are associated with molecular material of low excitation \citep{PlanckColdClumps}.
While the majority of these objects are not ``dark'' in a dynamical sense, they demonstrate that cold and weakly emitting molecular phases are widespread and may contribute non--negligibly to the total gas mass.

Taken together, these observations firmly establish the existence of a significant reservoir of molecular gas that is not traced by CO emission.
Whether this component is dynamically dominant remains an open question; however, its presence is now well supported observationally.
In this context, the phenomenology explored by Pfenniger \& Combes acquires renewed relevance: if a substantial fraction of the molecular gas is both cold and highly structured, its contribution to galactic dynamics and disk stability may be more important than commonly assumed in standard halo--based models.
The results presented in this work should therefore be viewed as complementary dynamical diagnostics aimed at probing the possible role of such a component in disk galaxies.

}
%%%%%%%%%%%%%%%%%%%%%%%%%%%%%%%%%%%%%%%%
%%%%%%%%%%%%%%%%%%%%%%%%%%%%%%%%%%%%%%%%
%%%%%%%%%%%%%%%%%%%%%%%%%%%%%%%%%%%%%%%%

\section{Conclusions}
\label{sect:concl} 
We have determined the generalized rotation curves $v_c(R,z)$, and vertical, $a_z(R,z)$, accelerations of the MW 
using \emph{Gaia}~DR3 data in the range of Galactocentric radii $8.5 < R < 14~\mathrm{kpc}$ 
and vertical distances $-2 < z < 2~\mathrm{kpc}$. 
The analysis assumes that the Galaxy is in a steady-state equilibrium described by the Jeans equations, 
which imply that matter moves in the mean gravitational potential of the system. 
We have then considered two distinct mass models to fit the measured $v_c(R,z)$ and $a_z(R,z)$. 

The first is the \emph{standard halo model}, in which the stellar and gaseous components --- whose structural 
properties are assumed to be known with reasonable accuracy --- reside in a disk and move on quasi-circular orbits, 
while an additional spherical component is present at small galactocentric radii (the bulge). 
In this framework, the dominant DM component is dynamically cold, characterized by a nearly isotropic 
velocity dispersion, and is in equilibrium within a self-consistent, quasi-spherical gravitational potential.

The second is the \emph{DM disk model}, in which the stellar, gaseous, and DM components 
all belong to a common, flattened system and move within the same mean-field potential of the disk. 
In both models, the stellar and gaseous components share the same circular velocity, being in equilibrium with 
the total potential of the system. 
In the former case, this potential corresponds to the combined disk+halo distribution, whereas in the latter it 
is entirely associated with the disk itself.  In this second scenario, the DM component also shares the same circular velocity as the stellar and 
gaseous components, since all contribute coherently to the gravitational potential of the disk. 
This is, of course, a mean-field description that does not account for the possible differences in the dynamical 
origins or evolutionary histories of the various disk populations, such as young and old stars or halo stars.
Such a mean-field description is commonly adopted when constructing mass models of the Galactic disk 
that include multiple components --- such as the thin, thick, and neutral hydrogen disks ---  to fit the observed 
rotation curve \citep[e.g.][]{Eilers_etal_2019,SylosLabini_etal_2023,Ou_etal_2024}.

Another simplifying assumption common to both models is that all disk components are described by a 
double-exponential density profile.  Although deviations from this idealized form may occur, for instance due to the flaring that characterizes 
the outer regions of the Galactic disk \citep{Chrobakova_etal_2022,SylosLabini_2024}, this approximation remains fully adequate within the limited range  of radial and vertical distances considered in this analysis.

Our results show that the DM disk model provides a significantly better agreement with the data 
than the standard Navarro–Frenk–White (NFW) halo profile. 
In particular, spherical halos with characteristic scale radii of order 
$\sim10\,\mathrm{kpc}$ contribute only marginally to the off-plane rotation curve and to the 
vertical acceleration within the inner disk, leaving these quantities predominantly determined 
by the distribution of the stellar mass. 
As a consequence, halo-based models systematically underestimate the measured vertical accelerations 
and fail to reproduce the observed decline of the rotation curve at intermediate heights. 
In contrast, models in which the DM is confined to a flattened, disk-like configuration 
predict substantial contributions to both the radial and vertical components of the gravitational field, 
leading to a markedly improved agreement with the observed trends of $v_c(R,z)$ and $a_z(R,z)$. 
This improvement is particularly evident at low to intermediate heights ($|z|\lesssim2\,\mathrm{kpc}$), 
where the vertical acceleration inferred from the data cannot be explained by the baryonic components alone. 

The success of the DM disk model arises from its geometry: 
a flattened mass distribution naturally enhances the vertical component of the gravitational potential 
without requiring an excessive total mass, and simultaneously reproduces the modest decline of the 
circular velocity with increasing $z$. 
These results strongly suggest that a significant fraction of the MW's dark matter is distributed 
in a disk-like structure rather than in a quasi-spherical halo. 
Forthcoming \emph{Gaia} data releases, offering improved statistics and reduced systematic uncertainties in 
stellar kinematics, will enable more stringent and spatially extended tests of the geometry of the 
Galaxy'ss dark matter component, potentially allowing one to constrain its vertical and radial scale lengths 
with unprecedented precision.

\section*{Acknowledgements}
We are grateful to Antonio Tedesco for discussions. 
F.S.L. is grateful to Michael Joyce  for many insightful and stimulating discussions.

This work presents results from the European Space Agency (ESA) space mission Gaia. Gaia data are being processed by the Gaia Data Processing and Analysis Consortium (DPAC). Funding for the DPAC is provided by national institutions, in particular the institutions participating in the Gaia MultiLateral Agreement (MLA). The Gaia mission website is https://www.cosmos.esa.int/gaia. The Gaia archive website is https://archives.esac.esa.int/gaia.

%%%%%%%%%%%%%%%%%%%%%%%%%%%%%%%%%%%%%%%%%%%%%%%%%%%%%%%%%%%%%%

%%%%%%%%%%%%%%%%%%%%%%%%%%%%%%%%%%%%%%%%%%%%%%%%%

\appendix
\section*{A Poisson Solver} 
\label{sect:poisson}

We solve the axisymmetric Poisson equation for the gravitational potential \(\Phi(R,z)\) in cylindrical coordinates:
\begin{equation}
\label{eq:poisson}
\frac{1}{R}\frac{\partial}{\partial R}\!\left(R\,\frac{\partial \Phi}{\partial R}\right)
+\frac{\partial^2 \Phi}{\partial z^2}
= 4\pi G\,\rho(R,z),
\end{equation}
where \(G\) is the gravitational constant and \(\rho\) is the mass density. The solver assumes axial symmetry (\(\partial/\partial\phi=0\)) and computes derived circular speeds
\begin{equation}
\label{eq:vc}
v_c(R,z) \equiv \sqrt{\,R\,\frac{\partial \Phi}{\partial R}\,}.
\end{equation}

%\paragraph{Units.}
The implementation uses astrophysical units with
\(\,G={4.30091 \times 10^{-6}}\ \mathrm{kpc}\,(\mathrm{km/s})^2/\mathrm{M}_\odot\).
Grid coordinates \(R\) and \(z\) are in kpc, \(\rho\) in \(\mathrm{M}_\odot/\mathrm{kpc}^3\), \(\Phi\) in \((\mathrm{km/s})^2\), and \(v_c\) in \(\mathrm{km/s}\).

%\section{Domain, Grid, and Discretization}
We discretize \([0,R_{\max}]\times[z_{\min},z_{\max}]\) on a uniform grid:
\[
R_i=(i-1)\,\Delta R\quad (i=1,\dots,N_R),\qquad
\]
\[
z_j=z_{\min}+(j-1)\,\Delta z\quad (j=1,\dots,N_z).
\]
A finite-volume/finite-difference scheme is used for the radial operator to preserve the cylindrical divergence form in Eq.\ref{eq:poisson}. 
At an interior point \((i,j)\),
\begin{flushleft}
\begin{align}
\frac{1}{R_i}\frac{\partial}{\partial R}\!\left(R\,\frac{\partial \Phi}{\partial R}\right)
&\approx \frac{1}{R_i \Delta R^2}\bigg[
\underbrace{\tfrac{R_i+R_{i+1}}{2}}_{R_{i+1/2}}\big(\Phi_{i+1,j}-\Phi_{i,j}\big) \notag \\[6pt]
&\quad -\underbrace{\tfrac{R_i+R_{i-1}}{2}}_{R_{i-1/2}}\big(\Phi_{i,j}-\Phi_{i-1,j}\big)
\bigg], \label{eq:disc}\\[6pt]
\frac{\partial^2 \Phi}{\partial z^2}
&\approx \frac{\Phi_{i,j+1}-2\Phi_{i,j}+\Phi_{i,j-1}}{\Delta z^2}.
\end{align}
\end{flushleft}
Defining
\begin{flushleft}
\begin{align}
R_{i\pm\frac12}\equiv \frac{R_i+R_{i\pm1}}{2},\qquad
\\[6pt]
a_{R}^{+}=\frac{R_{i+\frac12}}{R_i\Delta R^2},\quad
a_{R}^{-}=\frac{R_{i-\frac12}}{R_i\Delta R^2},\quad
a_z=\frac{1}{\Delta z^2},
\end{align}
\end{flushleft}
the discrete Laplacian at \((i,j)\) is
\bea
\label{eq:lapdisc}
&&
\mathcal{L}\Phi_{i,j} \equiv a_R^{+}\,\Phi_{i+1,j} + a_R^{-}\,\Phi_{i-1,j}
\\ \nonumber && + a_z\,\Phi_{i,j+1} + a_z\,\Phi_{i,j-1}
-\Big[(a_R^{+}+a_R^{-})+2a_z\Big]\Phi_{i,j}.
\eea
The right-hand side is \(4\pi G\,\rho_{i,j}\).

For the boundary conditions
\begin{itemize} \item {Axis} \(R=0\) (\(i=1\)): impose the regularity (Neumann) condition \(\partial \Phi/\partial R=0\).
Practically, \(\Phi_{1,j}=\Phi_{2,j}\).
\item {Outer radial boundary} \(R=R_{\max}\) and {vertical boundaries} \(z=z_{\min},z_{\max}\): default {Dirichlet} \(\Phi=0\), approximating vanishing potential at large distances.%
\end{itemize}

We solve the linear system with Successive Over-Relaxation (SOR). One SOR sweep updates interior points by
\bea
\label{eq:sor}
&& \Phi_{i,j}^{\text{new}}
= (1-\omega)\,\Phi_{i,j}^{\text{old}}
+
\\ \nonumber &&  \omega\,\frac{
a_R^{+}\,\Phi_{i+1,j}^{(*)} + a_R^{-}\,\Phi_{i-1,j}^{\text{new}}
+ a_z\big(\Phi_{i,j+1}^{(*)}+\Phi_{i,j-1}^{\text{new}}\big) - (4\pi G)\,\rho_{i,j}
}{
(a_R^{+}+a_R^{-})+2a_z},
\eea 
with relaxation parameter \(\omega\in(1,2)\). Symbols \((*)\) denote values already updated in the current sweep (Gauss-Seidel ordering). Convergence is monitored by the max-norm residual
\begin{equation}
\label{eq:res}
\mathcal{R}_{i,j} \equiv \mathcal{L}\Phi_{i,j} - (4\pi G)\,\rho_{i,j},\qquad
\|\mathcal{R}\|_{\infty}=\max_{i,j}|\mathcal{R}_{i,j}|,
\end{equation}
and the iteration stops when \(\|\mathcal{R}\|_{\infty}<\varepsilon\) (tolerance).

For the disk component the code provides, e.g.,  a double-exponential density (see Eq.\ref{eq:double_exp})
normalized so that the \emph{infinite} disk mass equals a user-provided \(M_{\rm d}\):
\bea
&&
\label{eq:rho0}
M_{\rm d} = \int_0^\infty\!\!\int_{-\infty}^{\infty} 2\pi R\,\rho\,dz\,dR
= 4\pi\,\rho_0\,R_{\rm d}^2\,z_{\rm d} \;.
\eea
Note in general the mass reported inside the finite box \([0,R_{\max}]\times[z_{\min},z_{\max}]\) will be smaller than \(M_{\rm d}\):
however if $z_{\max} = - z_{\min} \gg z_{\rm d}$ and $R_{\max} \gg R_{\rm d}$. 
After convergence, the circular speed is computed from Eq.\ref{eq:vc} using one-sided/forward differences near \(R=0\) and first-order differences elsewhere:
\begin{equation}
v_c(R_i,z_j) = \sqrt{\,R_i\,\frac{\Phi_{i,j}-\Phi_{i-1,j}}{\Delta R}\,},\qquad i\ge 2,
\end{equation}
with a forward difference for \(i=1\).

Concerning the convergence, the accuracy, and computational cost
\begin{itemize} 
\item {Relaxation \(\omega\)}: typical stable values are \(1.5\!-\!1.9\). If divergence occurs, reduce \(\omega\).
\item {Tolerance}: the default max-norm residual threshold \(\varepsilon\sim10^{-8}\) for \(\Phi\) in \((\mathrm{km/s})^2\) is usually sufficient for smooth models.
\item {Complexity}: one SOR sweep is \(\mathcal{O}(N_R N_z)\). Total cost is proportional to the number of iterations to convergence.
\item {Grid resolution}: decrease \(\Delta R,\Delta z\) for higher accuracy; ensure the domain \([0,R_{\max}]\times[z_{\min},z_{\max}]\) encloses the mass so that \(\Phi\approx 0\) at boundaries is reasonable. 
\end{itemize}

Improving the accuracy requires enlarging the integration domain and reducing the step sizes in both $R$ and $z$. However, this significantly increases the computational cost, which can become prohibitive on a normal workstation when exploring parameter space to determine the best–fit model by minimizing $\chi^2$ against observational data.

\end{document}